%% file: main.tex
\documentclass[numsec,webpdf,modern,medium]{oup-authoring-template}
\onecolumn % for one column layouts

\usepackage{siunitx}
\usepackage{booktabs}
\usepackage{subfig}
\newcommand{\anon}{1} % for anonymization
\usepackage{bbm}
\usepackage{booktabs}
\usepackage{longtable}
\usepackage{pdflscape}
\usepackage{subcaption}
\usepackage{graphicx}
\usepackage[nameinlink]{cleveref}
\crefname{section}{Section}{Sections}
\crefname{definition}{Definition}{Definitions}
\crefname{assumption}{Assumption}{Assumptions}
\crefname{theorem}{Theorem}{Theorems}
\crefname{lemma}{Lemma}{Lemmas}
\crefname{equation}{Equation}{Equations}
\crefname{figure}{Figure}{Figures}
\crefname{table}{Table}{Tables}
\crefname{appendix}{Appendix}{Appendices}

\graphicspath{{Fig/}}
\theoremstyle{thmstyleone}%
\newtheorem{theorem}{Theorem}%  meant for continuous numbers
\theoremstyle{thmstyletwo}%
\theoremstyle{thmstylethree}%
\newtheorem{definition}{Definition}
\newtheorem{assumption}{Assumption}
\newtheorem{lemma}{Lemma}

\newcommand{\R}{\mathbb{R}}

\makeatletter
\let\ps@opening\ps@empty
\newcommand*{\ps@rightnumbered}{%
  \ps@headings
  \let\@evenhead\@oddhead
}
\makeatother

\begin{document}
\pagestyle{rightnumbered}
% define a convenience command for the package mention
\ifnum\anon=1
  \newcommand{\pkgcite}{\citep{GNVBC2026}}   % final: normal citation
\else
  \newcommand{\pkgcite}{[Repository link withheld for double-blind review]} % review: anonymized text
\fi

% \journaltitle{Journal Title Here}
% \DOI{DOI added during production}
% \copyrightyear{YEAR}
% \pubyear{YEAR}
% \vol{XX}
% \issue{x}
% \access{Published: Date added during production}
% \appnotes{Paper}

\firstpage{1}

%\subtitle{Subject Section}

\title[Short Article Title]{Spatiotemporally Consistent Multivariate Bias Correction for Climate Projections via Nested Vine Copulas}

\author[1,2,3,$\ast$]{Theresa Meier}
\author[1,2]{Erwan Koch}
\author[1,2]{Valérie Chavez-Demoulin}
\author[1,2,3]{Thibault Vatter}
% \ORCID{}

\address[1]{\orgdiv{Faculty of Business and Economics (HEC)}, \orgname{University of Lausanne}, 
\orgaddress{\country{Switzerland}}}
\address[2]{\orgdiv{Expertise Center for Climate Extremes}, %, Faculty of Business and Economics -
% Faculty of Geosciences and Environment}, 
\orgname{University of Lausanne}, 
\orgaddress{\country{Switzerland}}}
\address[3]{\orgname{University of Applied Sciences and Arts Western Switzerland (HES-SO)},
\orgaddress{\state{Geneva}, \country{Switzerland}}}

\corresp[$\ast$]{Corresponding author. \href{email:theresa.meier.1@unil.ch}{theresa.meier.1@unil.ch}}

% \received{Date}{0}{Year}
% \revised{Date}{0}{Year}
% \accepted{Date}{0}{Year}

%\editor{Associate Editor: Name}

%\FMAUBIO{\color{black!20}\rule{37pt}{37pt}\color{black}}{\author{Author Name.} This is sample author biography text for journals which include this element in the front matter. The values provided in the optional argument are meant for sample purposes. There is no need to include the width and height of an image in the optional argument for live articles. Check the journal's online author guidelines to see if this element is required.}

%\FMAUBIO{\color{black!20}\rule{37pt}{37pt}\color{black}}{\author{Author Name.}  This is sample author biography text in the front matter. The values provided in the optional argument are meant for sample purposes. There is no need to include the width and height of an image in the optional argument for live articles. This is sample author biography text for journals which include this element in the front matter. The values provided in the optional argument are meant for sample purposes. There is no need to include the width and height of an image in the optional argument for live articles. Check the journal's online author guidelines to see if this element is required.}

%\abstract{
%\textbf{Motivation:} .\\
%\textbf{Results:} .\\
%\textbf{Availability:} .\\
%\textbf{Contact:} \href{name@email.com}{name@email.com}\\
%\textbf{Supplementary information:} Supplementary data are available at \textit{Journal Name}
%online.}

\abstract{ Climate models are essential for understanding large-scale climate dynamics and long-term climate change, yet they exhibit systematic biases when compared with historical observations. Existing multivariate bias correction (MBC) approaches do not explicitly handle spatiotemporal dependence. However, preserving both spatiotemporal and inter-variable consistency is essential for realistic climate dynamics and reliable regional impact assessments.
  To address this gap, we propose a novel MBC method called GN-VBC that uses generalised additive models (GAMs) to disentangle spatiotemporal deterministic effects from stochastic residuals. To model joint distributions and dependencies across variables and locations, we introduce nested vine copulas (NVCs), a hierarchical vine merging strategy. NVC in the context of MBC combines two dependence levels: (i) spatial dependence across locations, modelled separately for each variable, and (ii) inter-variable dependence modelled at a selected reference location, which links the spatial models into a coherent multivariate and spatial structure.
  An application to Switzerland shows improvements in preserving inter-variable, spatial and temporal dependence across a wide range of evaluation metrics.} 


% \keywords[Abbreviations]{abbreviation1, abbreviation2, abbreviation3, abbreviation4}

%  \otherabstract[Additional Abstract]{Use this element for elements such as Graphical abstract, Lay summary, Translated abstract etc. Que cum aut etum qui ium dolupta ssequia autati odis demporepe ad et es alit rem repudaerae min et volorum re volupta nobit volectur aut fuga.}

%  \otherabstract[Graphical Abstract]{\colorbox{black!20}{\hbox to 0.97\textwidth{\vbox to 50pt{}}}}

% \boxedtext{Key Messages}{
% \begin{itemize}
% \item Key boxed text here.
% \item Key boxed text here.
% \item Key boxed text here.
% \end{itemize}}

\maketitle
\thispagestyle{empty}
%\begin{epigraph}
%Epigraph text. Ximporem qui reperov idempedit modio. Bisto imagnatem quae aceptis
%nobitae quid eum rae adignis quias-sit vellacc uptatur sunt quis rentis eaquasit alia deliquam
%rec-to consed unt. Empor sum ratur ressimusdae. Nam fugiae.
%\source{Epigraph source}
%\end{epigraph}

\input{sections/01_intro}

\input{sections/02_gbc}

\input{sections/03_nvc}
\input{sections/04_gnvbc}

\input{sections/05_case_study}
\input{sections/06_discussion}

\section{Conflicts of interest}
The authors declare that they have no competing interests.

\section{Funding}
This research was supported in
part by the Hasler Stiftung under grant 2025-05-01-519.

\section{Data availability}
Both \href{https://cds.climate.copernicus.eu/datasets/projections-cordex-domains-single-levels?tab=overview}{CORDEX} and \href{https://cds.climate.copernicus.eu/datasets/reanalysis-era5-land?tab=overview}{ERA5-Land} data are publicly available in the Climate Data Store. MeteoSwiss TabsD and RhiresD \href{https://opendatadocs.meteoswiss.ch/c-climate-data/c3-ground-based-climate-data}{grid-data products} are publicly available. Their remapped version to the EUR-11 $0.11^{\circ}$ spatial resolution may be provided on request.

\section{Author contributions statement}

%Must include all authors, identified by initials, for example:
%S.R. and D.A. conceived the experiment(s),  S.R. conducted the experiment(s), S.R. and D.A. analysed the results.  S.R. and D.A. wrote and reviewed the manuscript.
T.M.: conceptualisation, data curation, formal analysis, investigation, methodology, project administration, software, validation, visualization, writing--original draft, writing--review and editing.\\
E.K.: conceptualisation, funding acquisition, methodology, resources, supervision, writing--original draft, writing--review and editing.\\
V.C.D.: conceptualisation, methodology, supervision, writing--original draft, writing--review and editing.\\
T.V.: conceptualisation, formal analysis, funding acquisition, investigation, methodology, software, supervision, writing--original draft, writing--review and editing.

\section{Acknowledgments}
The authors gratefully acknowledge Sven Kotlarski (MeteoSwiss) for providing Swiss data products,
the CORDEX initiative, as well as ECMWF for providing ERA5-Land data.

%UNCOMMENT THE BELOW TWO LINES IN CASE YOU NEED NUMBERED FORMAT.
%\bibliographystyle{oup-plain}
%\bibliography{reference}

%UNCOMMENT THE BELOW TWO LINES IN CASE YOU NEED AUTHOR YEAR FORMAT.
\bibliographystyle{oup-abbrvnat}
\bibliography{reference}

%%%%%%%%%%%%%%
\clearpage
\begin{appendices}
\input{sections/Appendix}

\end{appendices}

\end{document}

%% file: sections/01_intro.tex
\section{Introduction}\label{sec-intro}
% \begin{itemize}
%     \item Introduction to VBC (and QM) and the research gap they inherit
%     \item General motivation for CUVEE and GAM for bias correction
%     \item Definition of temporal and spatial consistency
% \end{itemize}

Global climate models (GCMs) provide physically consistent simulations of the Earth system and remain the foundation for understanding large-scale climate dynamics and long-term climate change \citep[e.g.,][]{flato2013evaluation}. Their coarse spatial resolution, however, limits their ability to capture regional processes and extremes. Regional climate models (RCMs) complement GCMs by dynamically downscaling large-scale climate information, adding finer spatial detail and a more realistic representation of mesoscale processes such as topography-driven precipitation and land-atmosphere feedbacks \citep[e.g.,][]{Rummukainen_2016}. Together, GCMs and RCMs provide complementary climate information that supports the assessment of changes in climate hazards and extreme events. Although the resulting projections remain too coarse for local‑scale impact studies, they constitute the essential basis for subsequent downscaling and impact‑oriented modelling frameworks used in adaptation planning. 

However, GCMs and RCMs exhibit biases, i.e.,~systematic differences between simulated climate statistics (such as a mean, quantile, or event frequency) and the corresponding statistics computed from observations or from a reference dataset \citep[e.g.,][]{Ehret2012}. For instance, temperature biases of up to $\pm1.5^\circ\mathrm{C}$ and precipitation biases reaching $\pm40\%$ have been reported over Europe \citep[e.g.,][]{kotlarski2014}, while multi-model ensembles often show systematic regional biases in precipitation, temperature, or wind \citep[e.g.,][]{vautard2021}. These biases arise from limited spatial resolution, simplified physical processes, and incomplete knowledge of the climate system, and may affect not only individual climate variables but also their joint behaviour across locations and time. This motivates the widespread use of \emph{bias correction} before applying climate model outputs in impact assessments and risk analyses. The general idea of bias correction is to identify discrepancies between a reference dataset, which is typically made up of observations or reanalysis data, and climate model outputs during the \emph{calibration period}, i.e.,~in a time period when both reference and model data are available. In a subsequent step, these identified patterns are used to modify projections for the future (\emph{projection period}). Here, a central challenge is to improve agreement with the reference climate without suppressing the climate-change signal represented by the simulations.

To formalise the notion of bias correction, let $Y$ denote a climate variable. We use superscripts $m$ and $r$ to refer to model and reference data, and $c$ and $p$ to denote the range of timestamps corresponding to the calibration and projection periods. For example, $Y^{(mc)}=\{Y^{(m)}_t:t\in c\}$ denotes model data during the calibration period. In our application, we select a projection period for which reference data are also available, allowing us to evaluate our approach. A bias correction is a transformation $h$ applied to model output such that the corrected values $\widetilde Y^{(m)}_t=h(Y^{(m)}_t)$ yield a projected series $\widetilde Y^{(mp)}$ whose distribution resembles that of the (unobserved) reference data during the projection period, $Y^{(rp)}$. In the univariate stationary setting, this reduces, e.g.,~to quantile mapping \citep[QM; e.g.,][]{rajczak2016robust,ivanov2017assessing}, where $h(\cdot)=\widehat Q^{(rc)}\circ\widehat F^{(mc)}(\cdot)$  with $\widehat{Q}^{(rc)}$ and $\widehat{F}^{(mc)}$ the empirical quantile and distribution functions of the reference and model data during the calibration period, respectively. 
% In practice, however, since climate processes are inherently multivariate and spatially structured, bias correction must address more than univariate discrepancies. 

%In particular, it must also preserve inter-variable dependencies, spatial consistency, and temporal dynamics. 

A wide range of bias correction methods has been proposed to enhance the fidelity of climate model outputs. Early univariate approaches relied on delta or scaling adjustments, which correct mean or variance biases while preserving modelled change signals. More flexible distribution-based approaches were later developed, most notably the aforementioned quantile mapping and its extensions such as detrended quantile mapping (DQM) and quantile delta mapping (QDM), which aim to preserve long-term climate trends while correcting distributional biases \citep{Cannon2015, rajczak2016robust,ivanov2017assessing}. These methods, however, operate on each variable and each location independently, therefore neglecting dependencies between variables, locations and time steps. Consequently, several multivariate bias correction (MBC) methods seek to correct biases in the joint distributions, ensuring that the resulting fields retain physically meaningful dependencies across variables, space, and time. State‑of‑the‑art approaches include correlation-based methods \citep{bardossy2012multiscale,cannon2016multivariate}, the normalization-rotation algorithm MBCn \citep{cannon2018multivariate}, the rank-resampling method R2D2 \citep{vrac2018multivariate}, and optimal-transport-based corrections such as dOTC \citep{robin2019}; see \cite{franccois2020multivariate} for a review.
More recently and closer to our work, the vine copula bias correction (VBC) framework \citep{funk2025} uses vine copulas to model complex dependence structures and the Rosenblatt transform \citep{rosenblatt1952remarks} for a multivariate analog of quantile mapping. Vine copulas are particularly attractive because they provide an explicit representation of complex dependence structures through collections of bivariate conditional copulas.
% VBC is applied location-wise and does not account for seasonality; limitations that we will address in our extension. 
% Nevertheless, most existing multivariate methods remain limited in their ability to simultaneously reproduce inter-variable, spatial, and temporal dependence. In addition, the underlying dependence structures are often not represented explicitly, and the three types of dependence are typically handled in the same way - if at all - which reduces flexibility and interpretability. 

Despite these developments, jointly correcting climate fields across multiple variables and locations remains challenging. Applying a multivariate correction independently at each location can improve local inter-variable dependence while leaving spatial relationships insufficiently controlled. Conversely, treating every variable-location combination as an unstructured multivariate vector may produce dependence structures that are difficult to interpret. Furthermore, the underlying dependence structures are often not represented explicitly, and the three types of dependence are typically handled in the same way, if at all, which reduces flexibility and interpretability. An additional challenge arises from systematic seasonal and spatial variation, which can complicate the estimation of stochastic dependence and affect the temporal properties of corrected series. These considerations motivate a framework that distinguishes systematic spatiotemporal variation from residual dependence, while exploiting the natural organization of climate fields into variables and locations.

% Beyond bias correction, vine copulas have also been used to model spatial dependence in environmental data. For example, \citet{graler2014} constructs distance-based vine models for spatial extremes, while \citet{erhardtczadoschepsmeier2015b} and \citet{Erhardt2015b} introduce vine structures parametrised by spatial covariates such as geographic distance and elevation. However, those approaches typically focus on a single variable across locations and therefore do not simultaneously address spatial and inter-variable dependence.

To this end, we propose GN-VBC, a multivariate bias correction method based on VBC that explicitly and simultaneously adjusts inter-variable, spatial, and temporal dependence. It combines two  complementary modelling steps. The first relies on generalised additive models (GAMs) to separate deterministic spatiotemporal effects from stochastic residuals. The second builds on nested vine copulas (NVCs), a hierarchical vine-merging framework that flexibly models joint distributions across variables and locations. The NVC construction combines two layers of dependence: (i) spatial dependence across locations for each variable, and (ii) inter-variable dependence at a selected reference location, represented through a multivariate vine structure linking the variable-specific spatial models into a coherent joint structure. The GAM layer is intended to ensure spatiotemporal consistency, while the NVC layer captures both spatial and inter-variable dependence. This approach has the advantage of being explicit, interpretable, and flexible enough to capture the three types of dependencies.

\begin{figure*}[t!]
  \centering
  \subfloat[\label{fig:map_CH}]{%
    \includegraphics[height=2in]{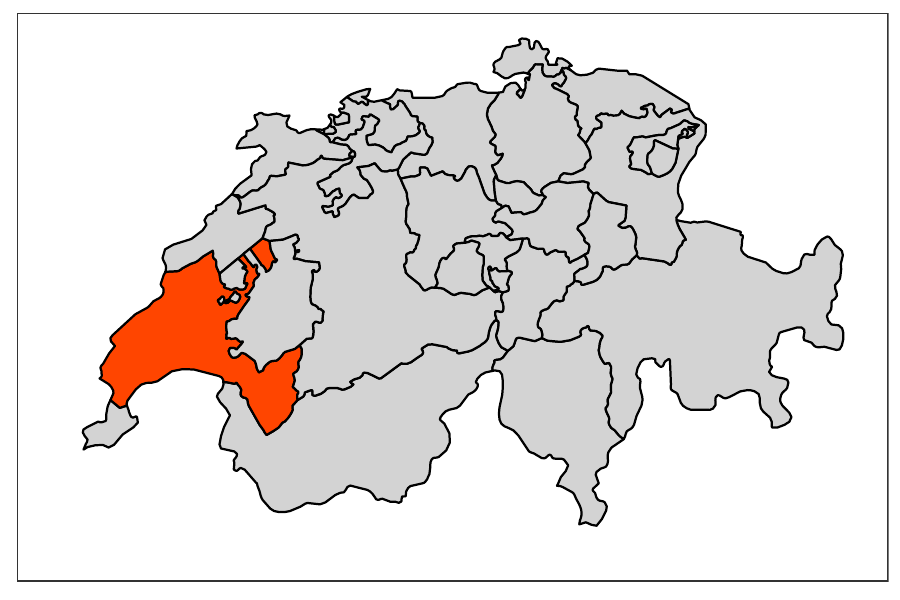}%
  }
  \hfill
  \subfloat[\label{fig:map_VD}]{%
    \includegraphics[height=2in]{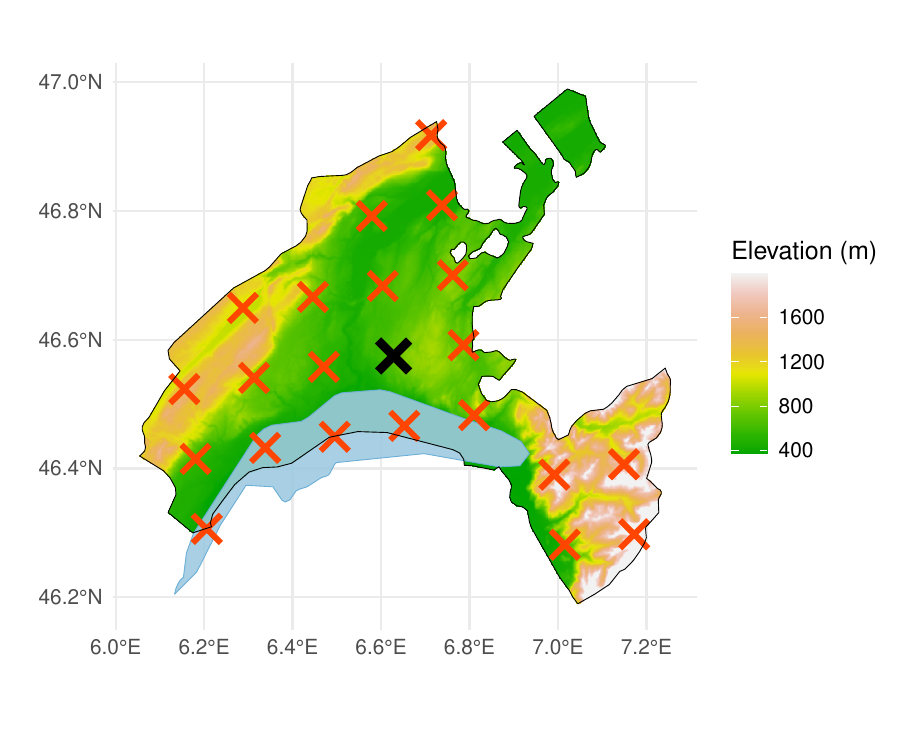}%
  }
  \caption{Panel (a) highlights the canton of Vaud in Switzerland. Panel (b) shows the 22 grid points in the study, with the black cell serving as the bridging location in \cref{sec:application}.}
  \label{fig:maps}
\end{figure*}

\begin{figure}[t!]
  \centering
  \includegraphics[width=\linewidth]{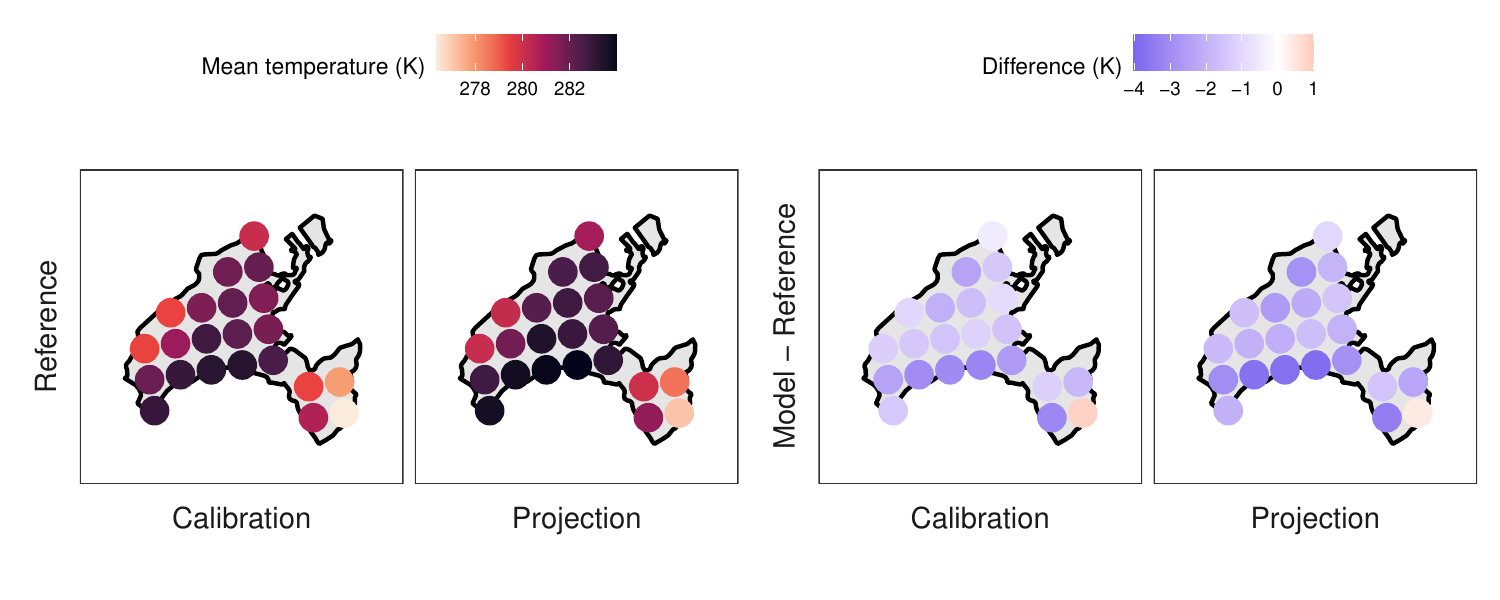}
  \caption{Mean reference temperature per grid point for the calibration (1980--2009) and projection (2010--2022) periods (left), and corresponding model-reference differences (right).}
  \label{fig:mean_tas}
\end{figure}

We apply our method to a challenging real-world setting in Switzerland (\cref{fig:maps}),
\begin{wraptable}{r}{7.5cm}
\vspace{-0.4cm}
  \centering
  \begin{tabular}{lll}
    \toprule
    Abbrev. & Atmospheric Variable\\
    \midrule
    \texttt{tas} & Mean 2-m temperature (\si{\kelvin}) \\
    \texttt{pr} & Total precipitation (\si{\milli\metre})\\
    \texttt{hurs} & Mean 2-m relative humidity \\
    \texttt{sfcWind} & Mean 10-m wind speed (\si{\metre\per\second}) \\
    \texttt{ps} & Mean surface pressure (\si{\pascal}) \\
    \bottomrule
  \end{tabular}
  \caption{Variables in the case study.}
  \label{tab:vars}
  \vspace{-0.2cm}
\end{wraptable}
where complex Alpine topography imposes strong spatial heterogeneity. The application involves jointly bias‑correcting the five key atmospheric variables of \cref{tab:vars}, projected within the Coordinated Regional Climate Downscaling Experiment for the European domain (EURO-CORDEX) at 22 grid points.
\autoref{fig:mean_tas} shows that the climate model (see \cref{sec:data} for a detailed description) underestimates mean temperatures across most locations. While this reflects a univariate bias (per variable and per location), further biases arise in inter‑variable and spatial dependence structures. Our objective in the application (\cref{sec:application}) is therefore to obtain bias‑corrected fields that faithfully reproduce reality in terms of inter-variable and spatiotemporal dependence.

We show that our method matches or outperforms state-of-the-art approaches, with potential operational implications for future Swiss climate projections. The Swiss Federal Office of Meteorology and Climatology (MeteoSwiss) regularly provides up-to-date national climate scenarios that are freely available for a wide range of applications. In the two most recent releases, CH2018 \citep{FISCHER2022100288} and CH2025, bias correction is performed using empirical QM, which does not account for inter‑variable, spatial, or temporal dependence. The methodology and results presented in this manuscript could therefore inform the next generation of Swiss climate scenarios and be directly relevant for future operational developments.

%The Vine Copula Bias Correction (VBC) \citep{funk2025} is a multivariate methodology designed to rectify systematic biases in climate model simulations while accounting for complex physical dependencies. The approach is anchored in vine copula theory, specifically generalised to accommodate variables with discrete-continuous mixtures, such as zero-inflated precipitation and radiation data

%Gap: we want to model the dependence between locations explicitly. By construction we account for. We want to explicitly model spatial dependency, temporal dependency and inter-variable dependency.

%By imposing this hierarchy, interpretable model. Graphical structure. Not black box like R2D2 (which is better suited to spatial dependence than MBCn).

% Post-processed climate outputs should
% meet several key
% requirements, including:
% \begin{itemize}
%     \item several features (e.g., change signal for
% the mean and extremes) should be
% conserved;
% \item inter-variable dependencies should be
% well preserved;
% \item extremes should be reliably represented
% at a very local scale;
% \item misrepresented relations in the original
% model output should be corrected;
% \item spatial consistency of the fields should
% be ensured;
% \item the temporal resolution should be
% sub-daily.
% \end{itemize}

The remainder of the paper is organised as follows. \cref{sec:gam,sec:nvc} present the GAM and NVC layers, respectively.
The full bias correction procedure is detailed in \cref{sec:gnvbc}. \cref{sec:application} presents the application to our Swiss case study, and \cref{sec:dis} concludes with a discussion.

%% file: sections/02_gbc.tex
\section{Generalized Additive Models for Bias Correction}\label{sec:gam}
A common strategy to address spatiotemporal dependence is to decompose the data into a systematic component and a stochastic remainder. In this context, \emph{generalised additive models} \citep[GAMs, see e.g.,][]{hastie1990generalized,wood2025generalized} allow for nonlinear effects of covariates meant to capture the spatiotemporal structure such as time, latitude, longitude, altitude, and interactions thereof.

Let $Y$ be a climate variable of interest, i.e.,~the bias correction target, and $\mathbf{X} \in \mathbb{R}^p$ be a vector of covariates.
GAMs extend linear models by allowing the conditional mean of the response $\mu(\mathbf{x}) = \mathbb{E}[Y \mid \mathbf{X} = \mathbf{x}]$ to depend on the covariates through a sum of additive components
\begin{align*}
  g\bigl(\mu(\mathbf{x}) \bigr)   = \sum_{j=1}^m f_j(\mathbf{x}),
\end{align*}
the $m$ additive terms being linear or smooth functions, e.g.,~defined via spline bases, and $g(\cdot)$ is a link function connecting the linear predictor to the response.
Inference is carried out by assuming that the response $Y \mid \mathbf{X} = \mathbf{x} \sim F_{\mu(\mathbf{x})}$, where $F_{\mu}$ denotes an exponential-family distribution with mean $\mu(\mathbf{x})$, leaving the dispersion parameter aside for simplicity.
This offers flexible modelling while maintaining the additive interpretability of predictor effects.

From a fitted model, one can then compute the \emph{probability integral transform} (PIT) $U~=~F_{\hat{\mu}(\mathbf{X})}(Y)$ to map the response to a uniform distribution on $[0,1]$.
For zero-inflated variables such as precipitation, a randomised PIT draws $U$ uniformly on $[0, F_{\hat{\mu}(\mathbf{X})}(0)]$ when $Y = 0$.
A GAM-based equivalent to the quantile mapping approach described in \cref{sec-intro} can then be written as
\begin{align}\label{eq:QM}
  \widetilde{Y} = F^{-1}_{\hat{\mu}^{(rc)}(\mathbf{X})} \circ F_{\hat{\mu}^{(mc)}(\mathbf{X})}(Y),
\end{align}
using the fitted GAMs for the reference and model data during the calibration period, respectively.
To account for model misspecification, one can then add another QM-like layer in the middle of the transformation.

However, even when applied to a single location and a single variable, this approach has two drawbacks.
First, it relies on GAM-based extrapolation to the projection period, i.e.,~the absolute time is needed as a covariate, instead of, say, time-of-year, time-of-day, or other covariates strictly meant to capture a ``deterministic'' temporal structure.
Second, and more importantly, it does not explicitly account for the modelled climate change signal.
Our GAM-equivalent of the quantile delta mapping of \citet{Cannon2015} can be written as
\begin{align}\label{eq:delta}
  \widetilde{Y} = F^{-1}_{\hat{\mu}^{(rc)}(\mathbf{X}) + \hat{\mu}^{(mp)}(\mathbf{X}) - \hat{\mu}^{(mc)}(\mathbf{X})} \circ F_{\hat{\mu}^{(mp)}(\mathbf{X})}(Y),
\end{align}
where the model data during the projection period is transformed to the uniform scale using the GAM fitted on the same period, thus avoiding extrapolation in time issues; the location parameter in the inverse CDF is adjusted by the difference between the GAMs fitted on the projection and calibration periods, to preserve the modelled climate change signal.

This approach can be applied independently to each variable, and forms the basis for our method.
By construction however, it does not explicitly account for inter-variable dependencies.
When considering a single location, one could add a vine bias correction \citep{funk2025} layer in the middle of the transformation.
In the next section, we propose a class of models that extends this idea to the multi-location setting.

%% file: sections/03_nvc.tex
\section{Nested Vine Copulas}\label{sec:nvc}

Copulas allow for flexible construction of a joint distribution with arbitrary one-dimensional margins \citep[see e.g.,][for textbook treatments]{nelsen2007introduction,joe2014}.
Mathematically, a \emph{copula} is a multivariate distribution with standard uniform margins.
From \citet{sklar1959fonctions}, for any multivariate distribution $F$ with univariate marginal distributions $F_{i}$ for $i= 1, \ldots, d $, there exists a copula $C$ such that
\begin{align*}
    F\left(y_1,\,\cdots,\,y_d \right) = C\left\{ F_{1}(y_1),\,\cdots,\,F_{d}(y_d) \right\} \text{for all } y\in\R^d,
\end{align*}
which is unique if the margins are continuous.
It implies that the joint density is the product between the marginal densities and the copula density $c = \partial^d C/\partial u_1 \cdots \partial u_d$, so that the joint log-likelihood is the sum of the marginal log-likelihoods and that of the copula.
This can be exploited in a two-step procedure by first estimating each of the margins separately and then the copula \citep{genest1995semiparametric, joe1996estimation}.

\subsection{Vine Copulas}\label{subsec:vines}
While standard copulas are well understood and have found many applications, their flexibility to model complex data structures is limited.
This is because, in most copulas, dependencies between all subsets of variables are described by the same parametrization, which is often too restrictive.
Popularised in \citet{aasczadofrigessibakken2009}, \emph{vine copulas} allow for a finer-grained modelling approach, and offer closed-form expressions for the required conditional distributions.
Following the seminal work of \citet{joe1996} and \citet{bedfordcooke2001, Bedford2002}, any $d$-variate copula density $c$ can be decomposed into a product of $d(d-1)/2$ bivariate (conditional) copula densities.
The order of conditioning in this decomposition can be organised using a graphical structure, called regular vine (R-vine), a sequence of spanning trees $\mathcal{V} = \{ \mathcal{T}_j \}_{j= 1}^{d-1}$, with $\mathcal{T}_j=(N_j, E_j)$ the tree, nodes and edges at level $j$. The trees are nested in a way such that the nodes in the first tree correspond to the variables themselves, and the edges in a tree $\mathcal{T}_j$ become the nodes of the next tree $\mathcal{T}_{j+1}$.
Such a construction is formalised in the following definition.
\begin{definition}[\citet{czado2019analyzing}]\label{def:rvine}
  A sequence of $d-1$ spanning trees is an R-vine if $N_1 = \{1, \dots, d\}$, $N_j = E_{j-1}$ for $j = 2, \dots, d-1$, and the \textbf{proximity condition} holds, i.e., if two nodes in $\mathcal{T}_{j+1}$ are joined by an edge, the corresponding edges in $\mathcal{T}_{j}$ share a common node.
\end{definition}
A vine copula identifies each edge of the sequence, its structure, with a bivariate copula density, called a \emph{pair-copula}, and the density then factorises as a product of these pair-copulas and the marginal densities.

% \begin{wrapfigure}{r}{0.3\textwidth}
%   \vspace{-1cm}
%   \centering
%   \tikzstyle{VineNode} = [ellipse, fill = white, draw = black, text = black, align = center, minimum height = 0.1cm, minimum width = 0.1cm]
%   \tikzstyle{DummyNode}  = [draw = none, fill = none, text = white]
%   \newcommand{\xshiftLabela}{1.2cm}
%   \newcommand{\yshiftLabela}{0.0cm}
%   \newcommand{\xshiftLabelb}{-0.6cm}
%   \newcommand{\yshiftLabelb}{-0.4cm}
%   \begin{tikzpicture} [every node/.style = VineNode, node distance =0.7cm]
%     \node (1){1}
%     node[DummyNode] (D1-2) [below of = 1]{}
%     node  (2)   [below of = D1-2]{2}
%     node[DummyNode] (D2-3) [below of = 2]{}
%     node  (3)   [below of = D2-3]{3}
%     node (1-2)  [right of = D1-2, xshift = \xshiftLabela, yshift = \yshiftLabela]{1,2}
%     node[DummyNode] (D1-3_2) [right of = 2, xshift = \xshiftLabela, yshift = \yshiftLabela]{}
%     node  (2-3)   [below of = D1-3_2]{2,3};
%     \draw (1) to node[draw=none, fill = none, font = \footnotesize,
%     above, xshift = \xshiftLabelb, yshift = \yshiftLabelb] {1,2} (2);
%     \draw (2) to node[draw=none, fill = none, font = \footnotesize,
%     above, xshift = \xshiftLabelb, yshift = \yshiftLabelb] {2,3} (3);
%     \draw (1-2) to node[draw=none, fill = none, font = \footnotesize,
%     above, xshift = \xshiftLabelb, yshift = \yshiftLabelb] {1,3$\vert$2} (2-3);
%   \end{tikzpicture}
%   % \includegraphics[width = 0.6\textwidth]{figures/vine-tree}
%   \caption{A 3d vine.}
%   \label{fig:ex_tree}
% \end{wrapfigure}
\begin{wrapfigure}{r}{0.25\textwidth}
  \centering
  \includegraphics[width=0.9\linewidth]{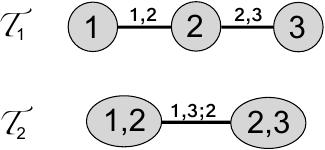}
  \caption{A 3d vine.}
  \label{fig:ex_tree}
\end{wrapfigure}

\cref{fig:ex_tree} shows an example with three variables.
In tree $\mathcal{T}_1$, variables 1 and 2 as well as variables 2 and 3 are linked by a pair-copula, the edge representing their connection serves as node in tree $\mathcal{T}_2$, and the corresponding copula density can be written as
\begin{align*}
c(u_1, u_2, u_3) &= c_{12}(u_1, u_2) \cdot c_{23}(u_2, u_3)  \cdot c_{13|2}(C_{1|2}(u_1|u_2),C_{3|2}(u_3|u_2)),
\end{align*}
where $C_{1|2}$ and $C_{3|2}$ denote the conditional distribution functions of $U_1$ and $U_3$ given $U_2$, respectively.

A vine-based multivariate distribution on $d$ variables can then be represented by the triplet
\begin{align}\label{eq:vine_copula}
  F := \left(\{F_j\}_{j=1}^d, \mathcal{V}, \mathbf{\mathcal{C}}(\mathcal{V})\right),
\end{align}
where $\{F_j\}_{j=1}^d$ denotes the marginal distributions, $\mathcal{V}$ a structure satisfying \cref{def:rvine}, and $\mathbf{\mathcal{C}}(\mathcal{V})$ is the corresponding collection of pair-copulas.
Because each margin and pair-copula can be specified independently, vine copula models are highly flexible and can be tailored to the specific needs of the application.
We refer to recent books and surveys \citep{joe2014,Aas2016,czado2019analyzing,czado2022vine} for comprehensive overviews.

\subsection{Nested Vine Copulas}\label{subsec:CUVEE}
In the spatial context, it is common to have separate models for different locations, and the question arises of how to combine them into a single model that captures the joint distribution across all locations.
Or one could envision a situation where one has separate models for different variables, and wants to combine them into a single model that captures the joint distribution across all variables.
In this section, we propose a new construction, the nested vine copula (NVC), which provides a framework for combining two separate vine-based multivariate distributions in a hierarchical manner.
While we illustrate the NVC for two vine copulas, the concept naturally generalises to $d \ge 2$ multivariate distributions.

Letting $\mathbf{Y}^{(1)} \in \mathbb{R}^{d_1}$ and $\mathbf{Y}^{(2)} \in \mathbb{R}^{d_2}$ be two (possibly dependent) random vectors, where $d_1$ and $d_2$ may differ, suppose that they follow a vine-based distribution as in \cref{eq:vine_copula}, denoted $F^{(1)}$ and $F^{(2)}$, respectively.
NVC derives another vine-based distribution $F$ of $\mathbf{Y} = (\mathbf{Y}^{(1)}, \mathbf{Y}^{(2)})^\top \in \mathbb{R}^{d_1 + d_2}$ by merging the graph structures and associated pair-copulas of $F^{(1)}$ and $F^{(2)}$, and capturing both shared and distinct dependence patterns.
A hierarchical construction from separate vines has advantages over a new model for the pooled data. When related datasets are available (e.g.,~from different regions, time periods, or experimental settings), separate models can uncover context-specific dependence structures. Additionally, from a computational standpoint, merging existing vine structures reduces computational cost by reusing previously estimated components.

Since both $\mathbf{Y}^{(1)}$ and $\mathbf{Y}^{(2)}$ admit vine-based representations, the marginals of $F$ are naturally defined as the union of the marginals of $F^{(1)}$ and $F^{(2)}$.
To combine $\mathcal{V}^{(1)}$ and $\mathcal{V}^{(2)}$ into a valid vine structure $\mathcal{V}$, we construct $\mathcal{T}_j$ for $j = 1, \dots, d_1+d_2-1$ as follows:
\begin{enumerate}
  \item[(1)] Add all nodes and edges from $\mathcal{T}_j^{(1)}$ and $\mathcal{T}_j^{(2)}$, that is $N_j^{(1)} \cup N_j^{(2)}$ and $E_j^{(1)} \cup E_j^{(2)}$, where $N_j^{(1)} = E_{j-1}^{(1)} = \emptyset$ for $j > d_1$ if $d_1 < d_2$ and conversely if $d_2 < d_1$.
  \item[(2)] Add a set of bridging edges $E_j^{(b)}$ satisfying the proximity condition to the disconnected graph 
  \begin{align*}
      \left (N_j^{(1)} \cup N_j^{(2)} \cup  E_{j-1}^{(b)}, E_j^{(1)} \cup E_j^{(2)}\right ),
  \end{align*} 
  so that the resulting graph 
  \begin{align*}
      \left (N_j^{(1)} \cup N_j^{(2)} \cup  E_{j-1}^{(b)}, E_j^{(1)} \cup E_j^{(2)} \cup E_j^{(b)}\right )
  \end{align*} is a spanning tree, starting with $E_0^{(b)}\equiv \emptyset$.
\end{enumerate}
The merged vine structure $\mathcal V$ thus combines the original dependence structures with additional cross-dependencies induced by the bridging edges. And we refer to \cref{app:math} in the supplementary material for a theorem on the validity of this procedure. The associated set of pair-copulas is given by
\begin{align*}
  \mathbf{\mathcal{C}}(\mathcal{V})=\mathbf{\mathcal{C}}(\mathcal{V}^{(1)})\cup\mathbf{\mathcal{C}}(\mathcal{V}^{(2)})\cup\mathbf{\mathcal{C}}(\mathcal{V}^{(b)}),
\end{align*}
where $\mathbf{\mathcal{C}}(\mathcal{V}^{(b)})$ denotes the pair-copulas corresponding to the bridging edges.
These additional copulas capture the dependence between nodes belonging to $\mathcal{V}^{(1)}$ and $\mathcal{V}^{(2)}$, ensuring that the resulting model defines a coherent joint distribution.

\begin{figure*}[t!]
  \centering
  \subfloat[\label{fig:example_merge}]{%
    \begin{minipage}[c][6cm][c]{0.51\linewidth}
      \centering
      \includegraphics[width=\linewidth]
        {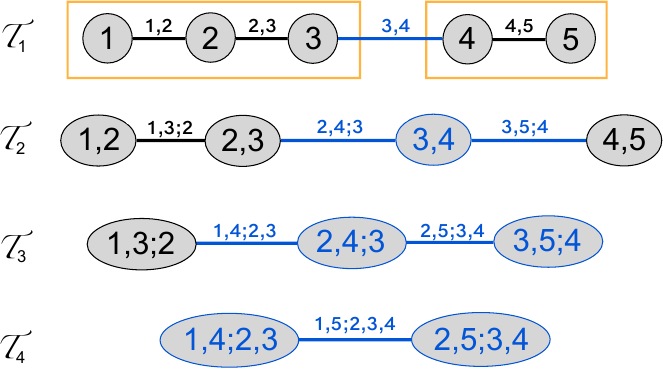}
    \end{minipage}
  }
  \hfill
  \subfloat[\label{fig:example_merge_options}]{%
    \begin{minipage}[c][6cm][c]{0.45\linewidth}
      \centering
      \includegraphics[width=\linewidth]
        {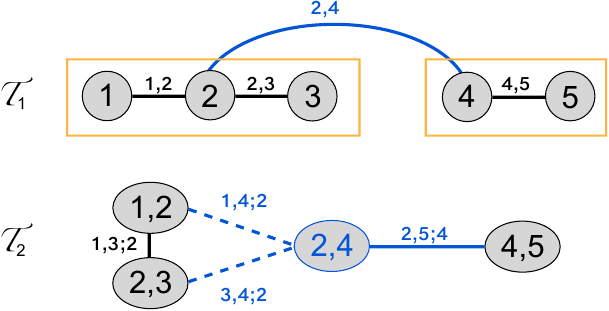}
    \end{minipage}
  }

  \caption{Merging two vines with node sets $N_1^{(1)} = \{1,2,3\}$ and $N_1^{(2)} = \{4,5\}$ with their edges as solid black lines and the bridging edges as solid blue lines. 
Figure (a) shows all trees with $\{3,4\}$ as first tree bridging pick. 
Figure (b) shows the first two trees with $\{2,4\}$ instead, with the dotted blue lines showing the two potential candidates to connect $\mathcal T_2$.}
  \label{fig:example_nvc}
\end{figure*}

\cref{fig:example_merge} illustrates this process. To initialise the recursive construction, a bridging edge $e \in \{j,k\}$ with $j \in \{1, 2, 3\}$ and $k \in \{4,5\}$ has to be selected to connect the nodes $N_1 = \{1, \dots,5 \}$ of $\mathcal{T}_1$.
Picking $e = \{3,4\}$ arbitrarily, the merged $\mathcal{T}_1$ has edge set $E_1^{(1)} \cup E_1^{(2)} \cup \{e\}= \{\{1,2\}, \{2,3\}, \{3,4\}, \{4,5\}\}$.
In $\mathcal{T}_2$, $e$ becomes part of the node set and appears in both bridging edges.
With this pick, only one configuration satisfies the proximity requirement, and the merging is uniquely determined afterwards.
Choosing $e = \{2,4\}$ instead, as shown in \cref{fig:example_merge_options}, the number of admissible
bridging edges in $\mathcal{T}_2$ increases and one needs to further pick either $\{1,4;2\}$ or $\{3,4;2\}$, in addition to $\{2,5;4\}$, to form $E_2^{(b)}$.
For $j \geq \max(d_1,d_2)=3$, the original vines have empty edge sets, so the merging procedure proceeds solely by adding bridging edges until $\mathcal{T}_{d_1+d_2-1} = \mathcal{T}_4$ is obtained.

We give details about the implementation in \cref{app:nvc-algorithm} of the supplementary material; we outline the main ideas here.
Both steps operate on tree-edge lists rather than the usual matrix representations of vine structures \citep{czado2019analyzing}, which avoids redundant re-encoding and keeps the merge independent of any particular array format.
Step~(1) is implemented as concatenation of edge lists: for each level $j$, the edges from $\mathcal{V}^{(1)}$ and $\mathcal{V}^{(2)}$ are merged into $E_j = E_j^{(1)}\cup E_j^{(2)}$.
Step~(2) proceeds level by level. At each level, admissible bridging edges are first enumerated by checking the proximity condition.
They are then ordered arbitrarily, manually, at random, or according to an optimality criterion, analogously to the vine selection algorithm of \cite{dissmann2013selecting}.
Finally, they are processed greedily in that order in a Kruskal-like fashion \citep{kruskal1956}, with acyclicity enforced via a disjoint-set union structure \citep{galler1964,tarjan1975}, until a spanning tree is reached.
The resulting merged vine is generally not unique and depends on the candidate-ordering rule.
With $n_j=d+1-j$ with $d=d_1 + d_2$ the number of nodes at level $j$, candidate generation scans $O(n_j^2)$ node pairs, each requiring set operations on supports of size $j$, giving a per-level bound of $O(n_j^2\,j)$.
Summing across levels gives an overall worst-case complexity bound that simplifies to $O(d^4)$.
In the structured setting of the next section however, the partially merged trees are nearly complete at low levels, and the routine is fast in practice for the dimensions considered.

\cref{app:sim:nvc} of the supplement contains experiments in a controlled setting to compare NVCs against standard vine copulas for various dimensions.

%% file: sections/04_gnvbc.tex
\section{Spatiotemporal Multivariate Bias Correction}\label{sec:gnvbc}
In this section, we describe \textbf{GN-VBC}, our spatiotemporal multivariate bias correction approach, incorporating \textbf{G}AMs and \textbf{N}VCs in the MBC method \textbf{VBC} \citep{funk2025}. VBC operates in three steps, on each location separately: First, in the \emph{estimation step}, vine copulas are estimated for both modelled projections and reference data in the calibration period. Then, in the \emph{correction step}, the multivariate distribution of the modelled projections is aligned with that of the reference data using the Rosenblatt transform \citep{rosenblatt1952remarks}, incorporating the simulated climate change signal between the calibration and projection periods (\emph{estimation step}). We adapt this procedure to model spatiotemporal dependencies and reformulate it in three conceptual steps:
\begin{enumerate}
  \item \textbf{Model}: the multivariate distributions of the reference data and the climate simulations, respectively in the calibration and projection periods.
  \item \textbf{Align}: the climate simulations with the distribution of the reference data in the calibration period using the estimated distributions.
  \item \textbf{Adjust}: the aligned data with the climate change signal, namely the discrepancy between the simulated evolution from the calibration to the projection period.
\end{enumerate}
Our GN-VBC method is implemented in the R package \texttt{GNVBC}, available on GitHub \pkgcite.
In \cref{app:sim_study_mbc} of the supplementary material, we compare GN-VBC to both QM and the VBC approach of \citet{funk2025}, illustrating how our method preserves spatiotemporal and multivariate relationships in a synthetic data setting.
The remainder of this section details each of the three steps.

\subsection{Modelling Step}\label{subsec:GAM_decomp}
\begin{figure}[t]
  \centering
  \begin{minipage}[c]{0.35\textwidth}
    \centering
    \subfloat[]{%
      \includegraphics[width=0.4\linewidth]{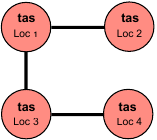}
    }\\[1em]
    \subfloat[]{%
      \includegraphics[width=\linewidth]{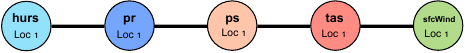}
    }
  \end{minipage}
  \hfill
  \begin{minipage}[c]{0.6\textwidth}
    \centering
    \subfloat[]{%
      \raisebox{-0.15\height}{%
        \includegraphics[width=\linewidth]{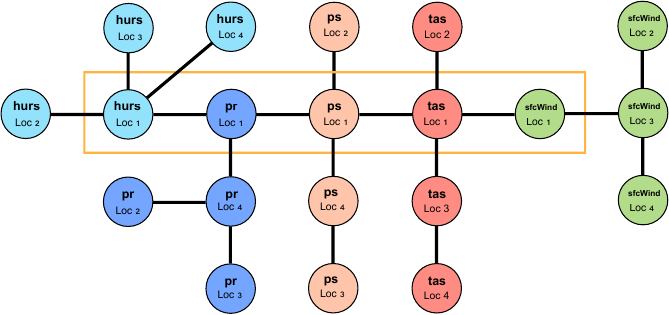}
      }
    }
  \end{minipage}
  \caption{Example of merging location- and variable-specific vine structures based on four locations and five variables in a $20$-dimensional vine copula. Panel~(a) shows the first tree of the location-specific vine for temperature; Panel~(b) the first tree of the variable-specific vine at location~1 which is highlighted in yellow in Panel~(c) which shows the resulting merged first tree. Variable abbreviations are given in \cref{tab:vars}.}
  \label{fig:loc-var}
\end{figure}

For $d$ variables and $s$ locations, let $Y^{(mp)} = \left \{Y_{i,j,t}^{(m)} : \, i = 1, \dots, d,\, j = 1, \dots, s, \, t \in p \right \}$ be the modelled projections, where $i$, $j$, and $t$ represent respectively the variable, location and time indices, and similarly for the modelled and reference data in the calibration period.
Treating each climate variable and dataset separately, we follow \cref{sec:gam} and fit a spatiotemporal GAM using a tensor-product spline in time and location, expressed as
\begin{align*}
    g_i(\mu_{i,j,t}) = f_i\big(d_t,\ \mathrm{Lat}_j,\ \mathrm{Lon}_j\big),
\end{align*}
where $\mu_{i,j,t}=\mathbb{E}[Y_{i}|\mathbf{X}=(t,j)]$, $d_t$ is the day-of-year at timestamp $t$, $\mathrm{Lat}_j$ and $\mathrm{Lon}_j$ are the latitude and longitude at location $j$, and we dropped the superscript for the sake of clarity.
For each dataset and variable, we then use the estimates to compute the PITs
\begin{align}\label{eq:PIT}
  U_{i,j,t} = F_i\!\left(Y_{i,j,t};\, \widehat{\mu}_{i,j,t}\right),
\end{align}
where $F_i(\cdot)$ denotes the cumulative distribution function of the fitted family for variable~$i$.
For each timestamp $t$, the PITs are then rearranged so that the values over all variables and locations form a single $ds$-dimensional vector
\begin{align*}
\mathbf{U}_t = \left(U_{1,1,t}, \dots, U_{1,s,t}, U_{2,1,t}, \dots, U_{d,s,t}\right) \in [0,1]^{ds}.
\end{align*}
Under the above ordering, the pair $(i,j)$ corresponds to column $(i-1)s+j$ of $\mathbf U_t$.
Extending \citet{funk2025} to the spatiotemporal context, one could then fit $ds$-dimensional unstructured vines separately to the samples $U^{(mp)} = \{\mathbf{U}_t^{(m)}: t \in p \}$ and $U^{(rc)}= \{\mathbf{U}_t^{(r)}: t \in c \}$.
Instead, we exploit the NVC of \cref{sec:nvc} and fit, for each PIT dataset,
\begin{itemize}
    \item a location-specific vine on the PITs with indices $\left\{(i,j)\right\}_{j=1}^s$ for each variable $i$,
    \item a variable-specific vine
    on the PITs with indices $\left\{(i, j_0)\right\}_{i=1}^d$ for a given $j_0$,
\end{itemize}
where $j_0$ is a bridging location.
For each PIT dataset, the $d$ location-specific vines are then merged with the NVC procedure into $ds$-dimensional vine-based distributions $\hat{F}^{(mp)}$ and $\hat{F}^{(rc)}$, with the variable-specific vine providing the bridging edges that determine how variables are connected across locations.
The two types of vines are complementary: the location-specific vines model within-variable spatial structure, while the variable-specific vine captures inter-variable dependence. An illustrative example is given in \cref{fig:loc-var}.

% An example for four hypothetical locations and the five variables used in the application study (see \cref{tab:vars} for their description) is visualised in \cref{fig:loc-var}. In the first step, we compute location-specific vine copulas for each variable; Panel~(a) shows the example for temperature. Each of the five variables yields a location-specific vine structure (only the first trees are shown). Using location~1 as the bridging location, a variable-specific vine copula on the five climate variables is estimated, as in Panel~(b). Finally, NVC merges the location-specific vine copulas according to the variable-specific edges, resulting in the merged first tree shown in Panel~(c). In this representation, the variable-specific vine determines the bridging edges $E_1^{(b)}$, highlighted by the yellow box.

\subsection{Alignment Step}\label{subsec:rosenblatt_transform}
As in \cite{funk2025}, we use the \emph{Rosenblatt transform} \citep{rosenblatt1952remarks} and its inverse as multivariate analogues of the PIT and the quantile function to bias-correct $U^{(mp)}$. For a distribution $F$ on $\R^d$, the transform and its inverse are maps $\mathcal{R}_F : \R^d \to [0,1]^d$ and $\mathcal{R}_F^{-1} : [0,1]^d \to \R^d$ defined recursively from a sequence of conditional distributions associated with $F$. If $\mathbf{Y}$ is a random vector with distribution $F$, then $\mathcal{R}_F(\mathbf{Y})$ is a random vector with independent uniform components. Conversely, $\mathcal{R}_F^{-1}$ maps a random vector with uniform components into one that has distribution $F$.
Although $\mathcal{R}_F$ and $\mathcal{R}_F^{-1}$ are generally unavailable in closed form, vine-based distributions are a notable exception.

It allows us to use the NVC-based $ds$-dimensional estimated distributions $\hat{F}^{(mp)}$ and $\hat{F}^{(rc)}$ to transfer the dependence structure of the reference calibration period to the biased model projections by successive applications of the Rosenblatt transform and its inverse, that is
\begin{align*}
\widetilde{\mathbf{U}}^{(m)}_t = \mathcal{R}_{\hat{F}^{(rc)}}^{-1} \circ \mathcal{R}_{\hat{F}^{(mp)}}\left(\mathbf{U}^{(m)}_t\right), \, t \in p,
\end{align*}
a multivariate analogue to the quantile mapping described in \cref{sec-intro}.
Using the projection-period vine copula model $\hat{F}^{(mp)}$ rather than $\hat{F}^{(mc)}$ avoids imposing stationarity of the dependence structure across periods and allows the bias correction to preserve the multivariate rank dynamics projected by the climate model. %By operating on PIT values rather than raw observations, GN-VBC avoids issues related to discreteness \citep{brockwell2007universal}.

\subsection{Adjustment Step}\label{subsec:mbc_step4}
To map the bias-corrected PITs $\widetilde{U}^{(mp)} = \left \{\widetilde{\mathbf{U}}^{(m)}_{t}: t \in p\right \}$ back to bias-corrected climate projections, we proceed as in \cref{eq:delta} in \cref{sec:gam} and define
\begin{align*}
\widetilde{Y}^{(m)}_{i,j,t}
=
F^{-1}_{\hat{\mu}^{(rc)}_{i,j,t} + \hat{\mu}^{(mp)}_{i,j,t} - \hat{\mu}^{(mc)}_{i,j,t}}
\left(
\widetilde{U}^{(m)}_{i,j,t}
\right),
\end{align*}
where $\hat{\mu}^{(rc)}_{i,j,t}$, $\hat{\mu}^{(mc)}_{i,j,t}$, and $\hat{\mu}^{(mp)}_{i,j,t}$ denote the fitted GAM means for the reference calibration, model calibration, and model projection datasets, respectively. This reverses the PIT step in \cref{eq:PIT} while preserving the modelled climate change signal through the additive correction $\hat{\mu}^{(mp)}_{i,j,t} - \hat{\mu}^{(mc)}_{i,j,t}$ in the location parameter.

%% file: sections/05_case_study.tex
\section{Application}\label{sec:application}

We apply the proposed methodology to the dataset introduced in \cref{sec:data}. The aim is to bias-correct daily simulations of five atmospheric variables (\cref{tab:vars}) at 22 grid points in the canton of Vaud (Switzerland). The calibration period is 1980–2009 and the projection period 2010–2022. Results compare the proposed GN-VBC pipeline to a set of benchmark methods summarised in \cref{tab:methods}.

As a univariate baseline, we use empirical QM. The multivariate methods considered fall into two categories: methods applied separately at each of the 22 locations (VBC and G-VBC) and methods applied jointly across all locations (all remaining methods).
\begin{table}[t]
  \centering
  \small
  \setlength{\tabcolsep}{6pt}
  \begin{tabular}{l|cc|ccc|cc|c}
    \toprule
    \textbf{Method}
    & \textbf{Uni.}
    & \textbf{Multi.}
    & \textbf{Indep.}
    & \textbf{Site-wise}
    & \textbf{Joint}
    & \textbf{Raw}
    & \textbf{PIT}
    & \textbf{NVC} \\
    \midrule 

    QM       & $\times$ &  & $\times$ &  &  & $\times$ &  &  \\ 
    MBCn     &  & $\times$ &  &  & $\times$ & $\times$ &  &  \\ 
    R2D2     &  & $\times$ &  &  & $\times$ & $\times$ &  &  \\ 
    VBC      &  & $\times$ &  & $\times$ &  & $\times$ &  &  \\ 
    C-VBC    &  & $\times$ &  &  & $\times$ & $\times$ &  &  \\ 
    G-VBC    &  & $\times$ &  & $\times$ &  &  & $\times$ &  \\ 
    GC-VBC   &  & $\times$ &  &  & $\times$ &  & $\times$ &  \\ 
    N-VBC    &  & $\times$ &  &  & $\times$ & $\times$ &  & $\times$ \\ 
    GN-VBC   &  & $\times$ &  &  & $\times$ &  & $\times$ & $\times$ \\   
    \bottomrule
  \end{tabular}
  \vspace{1em}
  \caption{Overview of bias correction methods. Columns 2 and 3 indicate whether the method is univariate or multivariate. Columns 4, 5 and 6 show how the method is applied: independently to each location and variable, site-wise, or jointly to all location-variable pairs. Columns 7 and 8 indicate whether raw data or PIT residuals are used within the method, and column 9 whether our vine merging algorithm, NVC, is incorporated.}
  \label{tab:methods}
\end{table}
In the first category, VBC is used either directly on the raw time series or on the probability integral transform (PIT) values obtained from GAM decompositions (G-VBC) based on a smooth effect for the day of the year only.
The second category includes two state-of-the-art MBC algorithms: MBCn proposed by \cite{cannon2018multivariate}, as well as R2D2 proposed by \cite{vrac2018multivariate}. In their work, \cite{cannon2018multivariate} applies MBCn  site-wise, while we follow \cite{franccois2020multivariate} who showcase its application to the joint dataset. For this study, we compared both performances (not shown) and opted for the joint application due to better results and, therefore, a stronger comparison with our proposed methods. For both MBCn and R2D2 we use the implementation in the \texttt{R} package \texttt{MBC}. Note that for R2D2, the package requires the same projection and reference period lengths, meaning that the calibration period narrows down to 1980-1992. This shorter calibration period may disadvantage R2D2, however, we retained the full calibration period for the other methods to avoid imposing a method-specific constraint on the entire comparison. In addition, we consider two naïve joint approaches based on standard VBC:  C-VBC, where VBC is applied to all variable-location pairs jointly using the raw time series, and GC-VBC, where the raw time series is replaced by PIT residuals. These approaches are considered naïve because the Dißmann algorithm (as implemented in \texttt{rvinecopulib} \citep{rvinecopulib} which is used in the implementation of \texttt{VBC}) is free to connect arbitrary variables across space, potentially leading to unintuitive pairings (e.g.,~humidity at one location paired with wind speed at a distant location).

This limitation motivates the N-VBC setting, where our NVC merging procedure constrains the vine structure to follow interpretable spatial and inter-variable relationships. Finally, GN-VBC denotes our entire proposed MBC pipeline as introduced in \cref{sec:gnvbc}, which applies our NVC merging algorithm to PIT values obtained from GAM decompositions based on a spatiotemporal tensor spline to estimate coherent nested vine structures across all variables and locations. This sequential setup isolates the contributions of the spatiotemporal GAM decomposition and the NVC merging step, allowing us to assess their effects independently. Note that in both G-VBC and GC-VBC, the delta mapping of VBC is replaced by our version of adjusting the mean structure (see \cref{subsec:mbc_step4}) not to account for varying climate conditions on the PIT scale.

An evaluation of the marginal model fit of our proposed GN-VBC pipeline is provided in \cref{app:margins} in the supplementary material.

The choice of GAM family in step~\ref{subsec:GAM_decomp} depends on the distributional characteristics of each variable. In this case study, across all methods using GAMs, we model temperature (\texttt{tas}) and surface pressure (\texttt{ps}) with a Gaussian distribution, wind speed (\texttt{sfcWind}) with a Gamma distribution, relative humidity (\texttt{hurs}) with a Beta distribution due to its support on $[0,1]$, and precipitation (\texttt{pr}) with a Tweedie distribution to account for the zero-inflated mixed discrete–continuous nature of rainfall.

The NVC merging uses a bridging location chosen near the canton centre (\cref{fig:map_VD}) to provide a stable anchor. To preserve geographic interpretability we optionally restrict the first tree of variable-specific vines to edges between adjacent grid points. Joint variable–location vines are truncated at level 12 for computational tractability; alternative truncation depths produced only minor changes in evaluation metrics.

Uncertainty is quantified with an annual block bootstrap \citep{kunsch1989jackknife}: entire years are resampled with replacement jointly across all variables and locations, thereby preserving within-year temporal and spatial dependence.

\cref{app:runtimes} of the supplementary material provides runtimes for all the methods considered in this application.

\subsection{Data}\label{sec:data}

We focus on five atmospheric variables (see \cref{tab:vars}) across the canton of Vaud in Switzerland (see \cref{fig:maps}) for the period 1 January 1980 to 31 December 2022, at daily temporal resolution and $0.11^\circ$ spatial resolution. Our reference data combine observational and reanalysis products. Specifically, temperature and precipitation are obtained from the TabsD and RhiresD grid-data products by MeteoSwiss \citep{MeteoSwissSpatialClimateAnalyses2021}.
%within Switzerland and extended with EOBS data outside Switzerland. 
Relative humidity, wind speed, and surface pressure are derived from ERA5-Land \citep[ECMWF Reanalysis 5th Generation for Land; see][]{hersbach2019era5} by averaging the hourly values. 

Climate simulations are sourced from the Coordinated Regional Climate Downscaling Experiment for the European domain (EURO-CORDEX) at the resolution of $0.11^\circ$ (EUR-11), using the RCP8.5 scenario \citep{giorgi2009addressing,jones2011coordinated}. The simulations are driven by the CNRM-CERFACS-CM5 global model and downscaled with the CLMcom-ETH-COSMO-crCLIM-v1-1 regional model over the EUR-11 domain. For consistency, both the MeteoSwiss observational datasets and ERA5-Land fields are remapped onto the common CORDEX grid using bilinear and conservative remapping, respectively \citep{mueller2010climate}. 

\subsection{Inter-variable Dependencies}\label{subsec:inter-var_con}
Our proposed GN-VBC construction first links each climate variable across locations before introducing bridges between the variable-specific vines. Consequently, variables observed at the same location are connected only through conditional dependence relationships involving other variable-location pairs. GN-VBC may therefore be at a slight disadvantage relative to methods that directly model location-wise inter-variable dependence, such as VBC; nevertheless, we demonstrate in this section that it achieves comparable performance in preserving these dependencies.

\begin{figure}[t]
  \centering
  \includegraphics[width=0.7\linewidth]{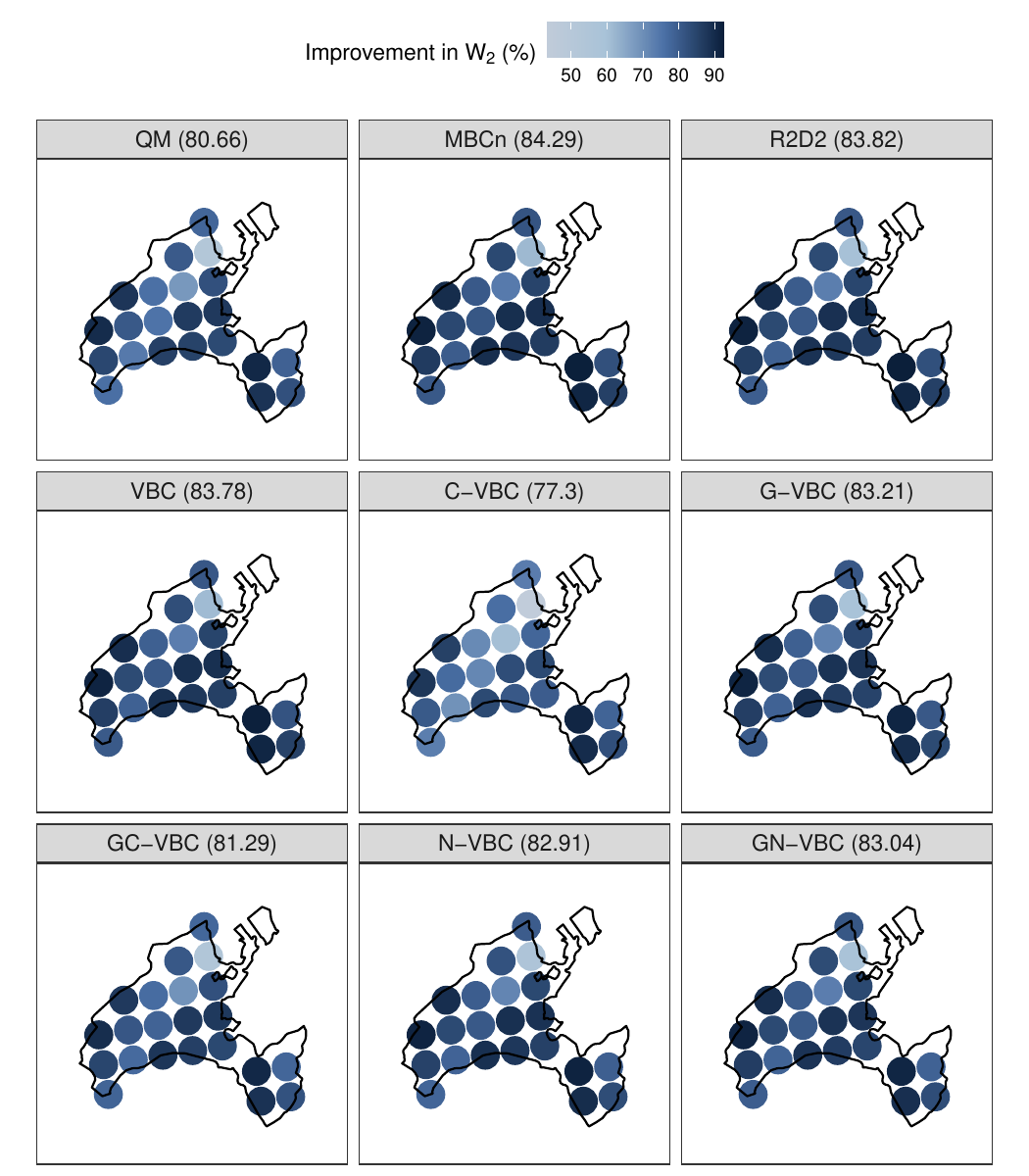}
  \caption{Improvement in multivariate second Wasserstein distances for five climate variables per location and BC method. The mean value over all locations is indicated in the title.}
  \label{fig:wd_loc}
\end{figure}

As in \cite{franccois2020multivariate} and \cite{funk2025}, we assess the preservation of inter-variable dependencies by comparing the reference and the bias-corrected distributions at each location using the Wasserstein distance \citep{villani2008optimal}.
For two distributions $F_1$ and $F_2$, its square is given by
$
  W_2^{2}(F_1, F_2)
  =
    \inf_{\gamma \in \Pi(F_1, F_2)}
    \int_{\mathbb{R}^d \times \mathbb{R}^d}
    \|x-y\|^2 \, d\gamma(x,y)$,
with $\Pi(F_1,F_2)$ the set of distributions on $\mathbb{R}^d \times \mathbb{R}^d$ with marginals $F_1$ and $F_2$, and $\|\cdot\|$ the Euclidean norm.
After computing 5-dimensional second Wasserstein distances over our climate variables for each location, we show in \cref{fig:wd_loc} the relative improvement
\begin{align}
\label{eq:WD_improv}
  \frac{
    W_2\!\left(F^{(rp)}, F^{(mp)}\right)
    -
    W_2\!\left(F^{(rp)}, \widetilde{F}^{(\mathrm{mp})}\right)
  }{
    W_2\!\left(F^{(rp)}, F^{(mp)}\right)
  },
\end{align}
where $\widetilde{F}^{(\mathrm{mp})}$ corresponds to the distribution of the bias-corrected model projections.
Please refer to \cref{app:tables} of the supplementary material for the exact values for each grid cell. Across all methods, a broadly consistent spatial pattern emerges, with substantial improvements typically between $70$ and $90\,\%$ for most grid points, and comparatively lower improvements in the northeastern part of the canton. Despite being a univariate method that does not explicitly model inter-variable dependence, QM's performance is only slightly worse than that of the established multivariate approaches.

Among the multivariate methods, R2D2, MBCn, and VBC yield similar levels of improvement. Incorporating the GAM-based decomposition prior to VBC (G-VBC) does not lead to an additional gain in this purely location-wise setting. In contrast, when locations are treated jointly, C-VBC, without structural constraints on the dependence across variables and locations, results in the lowest overall improvement. This performance loss is partially mitigated by incorporating the GAM decomposition in GC-VBC and more effectively addressed by introducing  nested vines in N-VBC, which enforces spatially coherent and interpretable vine structures. Combining both the GAM-based decomposition and the NVC merging algorithm yields improvements comparable to those achieved by established multivariate methods. The following subsections assess its ability to further preserve spatial and temporal consistency, the main objective of our method.

\subsection{Spatial Consistency}\label{subsec:spat_con}
\begin{figure}[t]
  \centering
  \includegraphics[width=\linewidth]{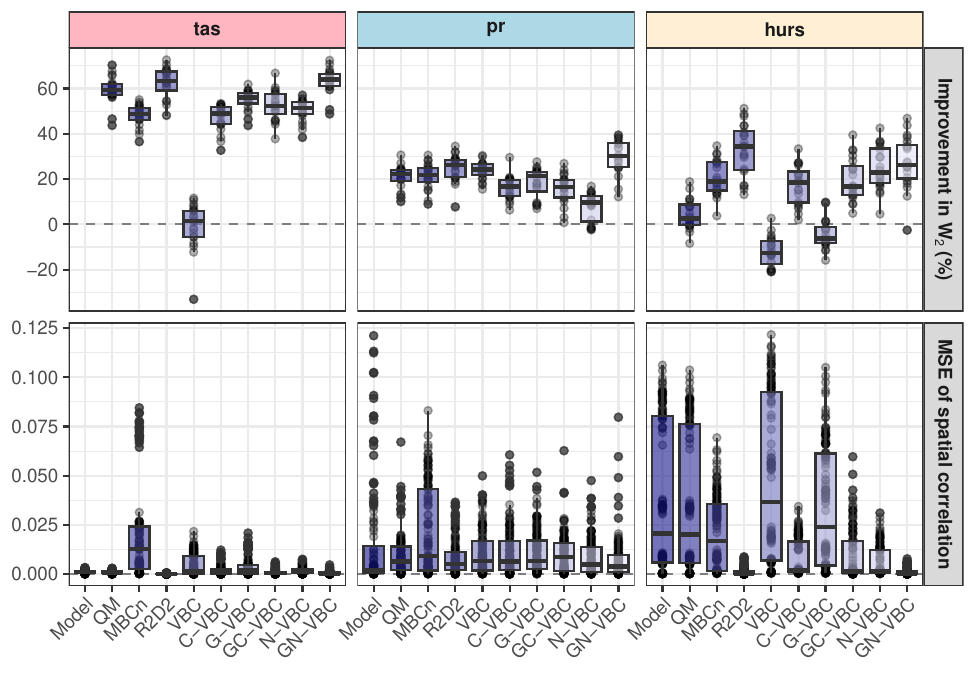}
  \caption{Evaluation of spatial consistency: improvement in second Wasserstein distance with respect to no correction  (top row, higher is better); MSE between the spatial correlation of the reference and the respective correction method (bottom row, lower is better).}
  \label{fig:spat}
\end{figure}
In our proposed framework, spatial coherence is promoted at two stages: first through the spatiotemporal decomposition via location-specific tensor splines, and second through the hierarchical merging of variable-specific vine copulas using NVC. To quantitatively assess how well spatial dependence is preserved, we follow \citet{franccois2020multivariate} and evaluate all methods using variable-specific second Wasserstein distances. Relative improvements with respect to the reference are computed as in \cref{eq:WD_improv}, but based on the joint $22$-dimensional distributions across all locations. The first row of \cref{fig:spat} shows the resulting improvements for three selected variables as boxplots over bootstrap replicates; numerical values for all variables are provided in \cref{app:tables} of the supplementary material.

For near-surface temperature (left column), median improvements range between approximately $48\%$ and $64\%$, with the notable exception of VBC applied independently at each location. Additional analyses (not shown) indicate that this poor performance is primarily driven by strong seasonal patterns in temperature. Since standard VBC operates directly on the raw time series and does not explicitly account for seasonality, vine estimation becomes unstable in this setting. This effect largely disappears when applying C-VBC, where the joint modelling of all variables and locations allows the vine copula to borrow strength across dimensions. Introducing the GAM-based decomposition prior to bias correction leads to a substantial improvement for both location-wise and joint applications of VBC (G-VBC and GC-VBC). Moreover, imposing a hierarchical structure via NVC slightly improves spatial consistency compared to unrestricted joint modelling. The strongest gains are achieved when combining GAM decomposition with NVC in GN-VBC. This suggests a complementary interaction between the two: the GAM decomposition stabilises marginal behavior by removing systematic spatiotemporal structure, while NVC enforces an interpretable spatial dependence structure.  R2D2, which has been shown to preserve spatial dependencies across multiple studies \citep{franccois2020multivariate, vrac2018multivariate}, performs comparably well to our approach, GN-VBC. As for MBCn, although it is applied to all locations simultaneously, it performs poorly.

A similar pattern is observed for relative humidity (right column), although improvements are generally more moderate than for temperature. Applying VBC independently at each location yields no improvements, and incorporating the GAM decomposition (G-VBC) leads to a noticeable enhancement, but the overall performance remains poor. This indicates that removing seasonal and smooth spatiotemporal effects alone is insufficient to recover realistic spatial dependence in relative humidity. A substantial improvement is achieved when VBC is applied jointly to all locations as in C-VBC. Moreover, adding the GAM decomposition to the joint VBC setup does not lead to marked additional gains. However, when introducing NVCs (as in N-VBC) or both components are combined (GN-VBC), just as discussed for temperature, a clear improvement emerges, with similar performance as R2D2. In contrast, empirical QM and MBCn show little to no improvement for relative humidity, reflecting their inability to address spatial dependence structures.

For precipitation (middle column), the differences between the correction methods are less pronounced, with median improvements ranging between approximately $9\%$ and $31\%$. Incorporating GAM decompositions (G-VBC and GC-VBC) slightly deteriorates the performance of both VBC and C-VBC. This suggests that, for precipitation, removing smooth spatiotemporal components is less beneficial. Introducing NVCs leads to a substantial loss in median performance over C-VBC, indicating that enforcing a structured and interpretable spatial dependence is not advantageous here. As observed for the other variables, the combination of GAM decomposition and NVC (GN-VBC) ultimately outperforms competing approaches, including R2D2, demonstrating that jointly addressing marginal stabilization and spatial dependence remains beneficial.

While Wasserstein distances quantify similarity between joint spatial distributions, a complementary indicator of spatial consistency is the preservation of spatial correlation between locations. Following \citet{largeau2025investigating}, we define
\[
B_r
=
\left\{
(s_1, s_2)\in\{1,\dots,s\}^2 :
d(s_1,s_2)
\in
\left[r-\frac{\Delta r}{2},\,r+\frac{\Delta r}{2}\right]
\right\},
\qquad
d_r = |B_r|.
\]
as the set of all pairs of locations whose pairwise distances fall within a distance bin centered at $r$ with radius~$\Delta r$. For standardised (in time) $\widetilde{Y}_{i,j,t}$ of length $N$, i.e.,~$t = 1, \dots, N$,  the spatial correlation function for variable $i$ is defined by
\[
\rho_{\widetilde{Y}_i}(r)
=
\frac{1}{N}
\sum_{t=1}^{N}
\frac{1}{d_r}
\sum_{(s_1,s_2)\in B_r}
\left(
\widetilde{Y}_{i,s_1,t}
-
\langle\widetilde{Y}_{i,t}\rangle
\right)
\left(
\widetilde{Y}_{i,s_2,t}
-
\langle\widetilde{Y}_{i,t}\rangle
\right)/ \text{std}_{\widetilde{Y}_i}^2,
\]
where for each time $t$ we define the spatial mean across all locations as
\[
\langle\widetilde{Y}_{i,t}\rangle
=
\frac{1}{s}
\sum_{j = 1}^s
\widetilde{Y}_{i,j,t} \quad \text{and} \quad \text{std}_{\widetilde{Y}_i} = \sqrt{\frac{1}{s}\sum_{j=1}^s \left(\widetilde{Y}_{i,j,t} - \langle\widetilde{Y}_{i,t}\rangle\right)^2.}
\]
In our setting, the distance bins are chosen to correspond to the length of the shortest path between location pairs when only adjacent grid cells are allowed to connect. This results in six distinct radii, with the largest radius representing pairs of locations connected through six adjacent edges.

For each variable and BC method, the spatial correlation function is evaluated at these six radii and compared to the reference spatial correlation function using the MSE. The results across bootstrap replicates are shown as boxplots in the second row of \cref{fig:spat}. For temperature, the raw model output already reproduces the spatial correlation structure reasonably well, leaving limited room for improvement. Consequently, most BC methods yield comparable performance, with the exception of MBCn and VBC, which exhibit higher variability in MSE values. For precipitation and relative humidity, the variability across bootstrap samples is substantially larger, reflecting the more heterogeneous spatial structure of these variables. GN-VBC shows a strong variance reduction across bootstrap replicates, indicating a more stable preservation of spatial correlation. In the case of relative humidity, VBC-based methods that model all locations jointly generally outperform location-wise corrections. In particular, C-VBC provides a clear improvement over standard VBC, highlighting the importance of explicitly modelling spatial dependence. Interestingly, the GAM decomposition alone hardly enhances the preservation of spatial correlations. However, when combined with the hierarchical vine structure imposed by NVC, it leads to similar performance as R2D2. This suggests that the benefits of removing spatiotemporal trends via GAMs are fully realised only when coupled with a structured multivariate dependence model. In contrast, QM does not improve upon the uncorrected model output for either precipitation or relative humidity, confirming that marginal corrections alone are insufficient to maintain spatial coherence.

\begin{figure}[t]
  \centering
  \includegraphics[width=\linewidth]{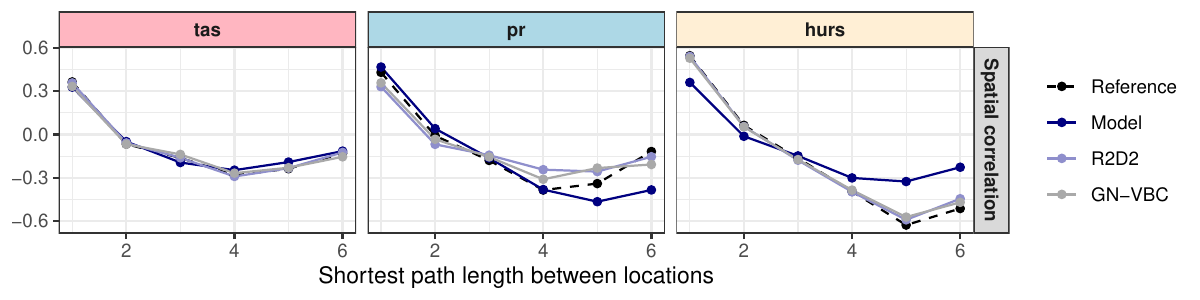}
  \caption{Spatial autocorrelation as a function over shortest path lengths between grid points for three variables. The dashed black line shows the reference. For clarity, only R2D2, and the proposed GN-VBC approach are shown.}
  \label{fig:spat_algo}
\end{figure}

To further illustrate the behavior of the spatial correlation function across the six distance classes considered in this study, \autoref{fig:spat_algo} shows that, as expected, across all variables spatial correlation generally decreases with increasing path length. For larger path lengths (five to six adjacency edges), a slight increase in correlation is observed. This effect arises from the elongated geometry of the canton of Vaud: due to its skewed shape (see \cref{fig:map_VD}), some location pairs that are geographically close in Euclidean distance are connected only through longer adjacency paths, leading to higher correlations at larger path lengths than for some intermediate distances.

While the differences between the model and the reference and bias-corrected temperature data are marginal, the differences between the model and the reference precipitation and relative humidity data are more pronounced at larger distances. Both R2D2 and GN-VBC perform equally well in improving the spatial correlation of the model, demonstrating that our novel MBC approach can compete with existing spatially coherent methods.

\subsection{Graphical Representation of the Dependence Structure}
\label{subsec:trees}

An attractive feature of vine copulas for estimating joint distributions is their graphical representation, which provides an interpretable factorization of the dependence structure. Through the hierarchical combination of spatial and inter-variable vines, the first tree of the estimated NVC provides insights into the dependence both across locations and between climate variables. This is particularly useful in bias correction, as it helps characterise how the variables co-vary across the study region.

Within our MBC framework, estimating an NVC for the modelled calibration data \(mc\) is not required. Nevertheless, such a model could also be estimated to provide an additional diagnostic comparison of the dependence structures in \(mc\), \(mp\), and \(rc\) (not shown). We emphasise that the displayed trees visualise the factorization of the joint copula density. They should not be interpreted as conditional-independence graphs in which the absence of an edge implies conditional independence.

\begin{figure*}
  \centering

  \subfloat[\label{fig:tree_mp}]{%
    \includegraphics[width=\textwidth]
      {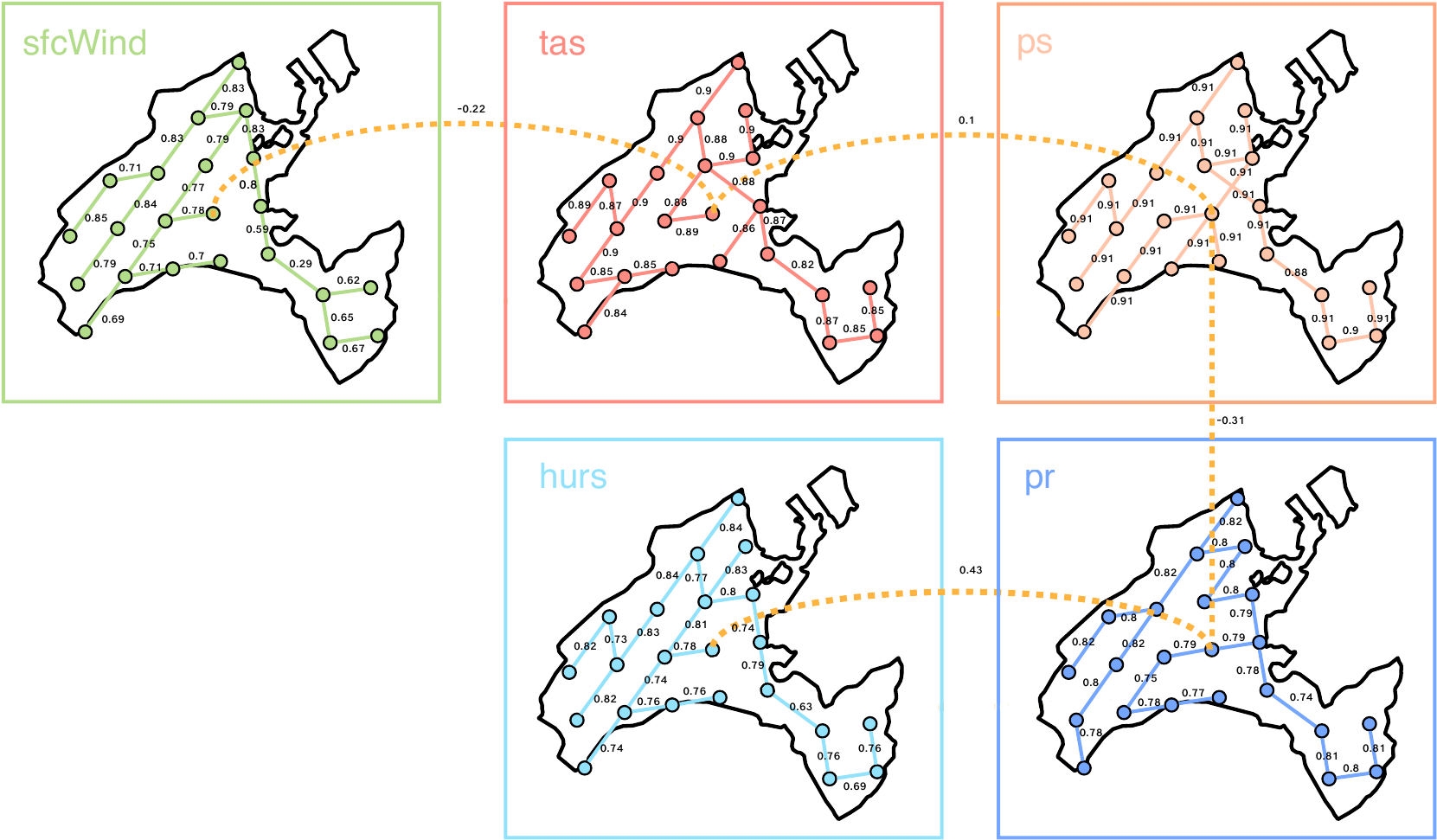}%
  }

  % \medskip

  \subfloat[\label{fig:tree_rc}]{%
    \includegraphics[width=\textwidth]
      {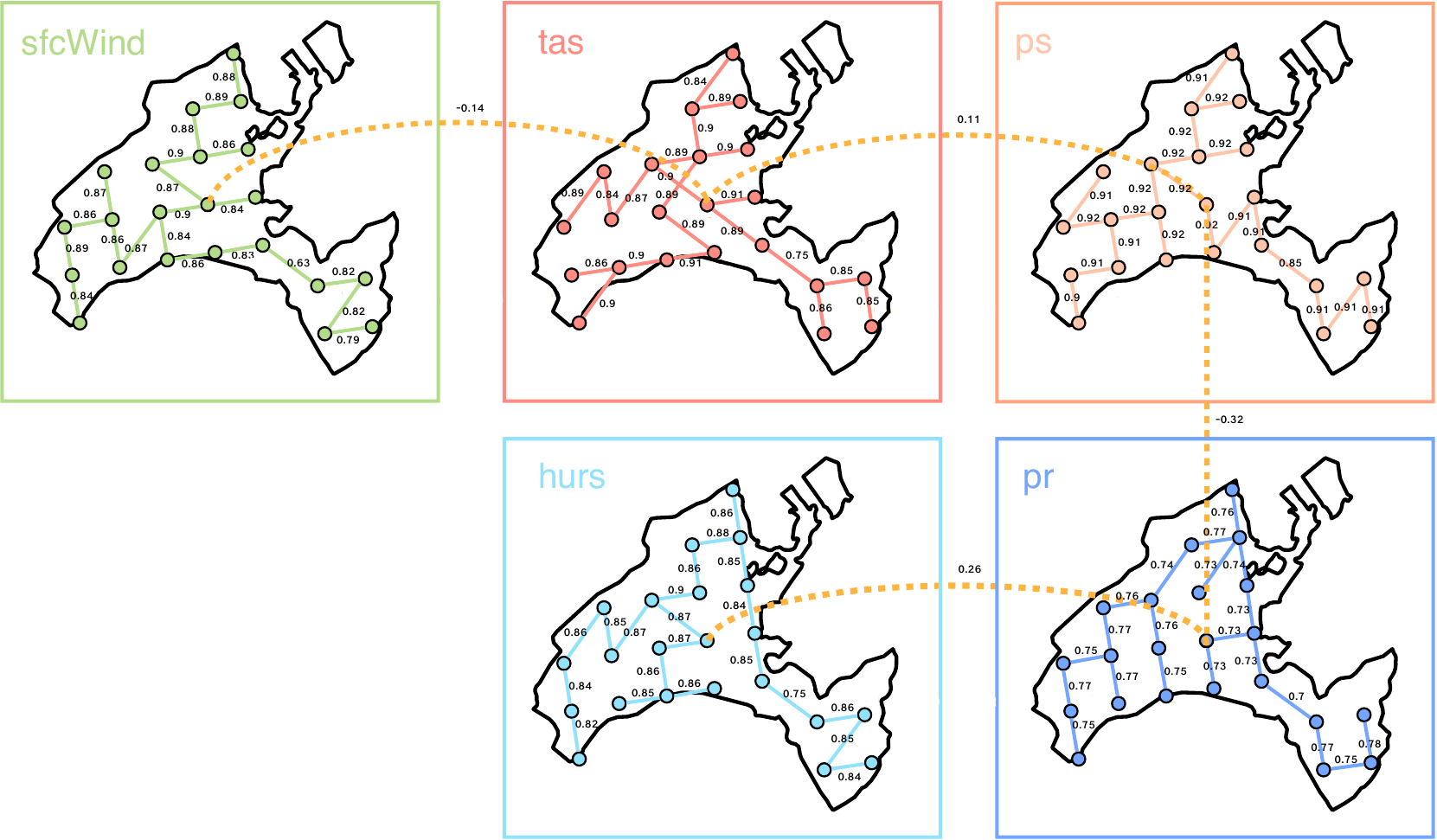}%
  }

  \caption{First trees of the hierarchical vine copula derived by the NVC merging algorithm for the model projection data $mp$ (a) and the reference calibration data $rc$ (b). Within-variable spatial edges are shown as solid lines, whereas dashed lines denote the bridging edges connecting the five spatial vines. The edge labels report the estimated Kendall's \(\tau\).}
  \label{fig:trees}
\end{figure*}

The first trees of the NVCs estimated during the modelling step of the proposed GN-VBC pipeline (see \cref{subsec:GAM_decomp}) shown in \autoref{fig:trees} reveal that, across both data sets, surface pressure and temperature exhibit the strongest and most spatially homogeneous dependence. Near-surface wind speed and, in particular, precipitation  display weaker and more heterogeneous spatial associations. A comparison between the model projections and reference calibration data indicates broadly similar dependence strengths for temperature and surface pressure, whereas more pronounced differences are apparent for wind speed and relative humidity.

The graphical representation also reveals geographically meaningful patterns. In the mountainous southeastern part of the canton (see \cref{fig:map_VD}), the spatial associations of wind speed are weaker than across much of the central and western study region, particularly in the model projections. This is meteorologically plausible because wind in complex Alpine terrain can be strongly affected by localised processes, including terrain-induced circulation, valley orientation, and orographic effects \citep{bugnard2025comparison}. More generally, this example illustrates how the fitted vine structure can expose spatial heterogeneity in the dependence that would be difficult to identify from a single aggregate measure.

Finally, the bridging edges connecting the five variable-specific spatial vines generally exhibit substantially weaker dependence than the within-variable spatial edges. This finding is consistent with the hierarchical construction adopted here, in which the spatial dependence within each variable is modelled before the variable-specific vines are connected through inter-variable edges.

\subsection{Temporal Consistency}\label{subsec:temp_con}

In the final step, we assess the proposed BC framework in terms of temporal consistency. Since the GAM decomposition explicitly models smooth seasonal and spatiotemporal effects through the day-of-year component, methods incorporating this step are expected to better preserve temporal dependence structures by construction.
\begin{figure*}[t!]
  \centering

  \subfloat[\label{fig:acf_365}]{%
    \includegraphics[width=\textwidth]
      {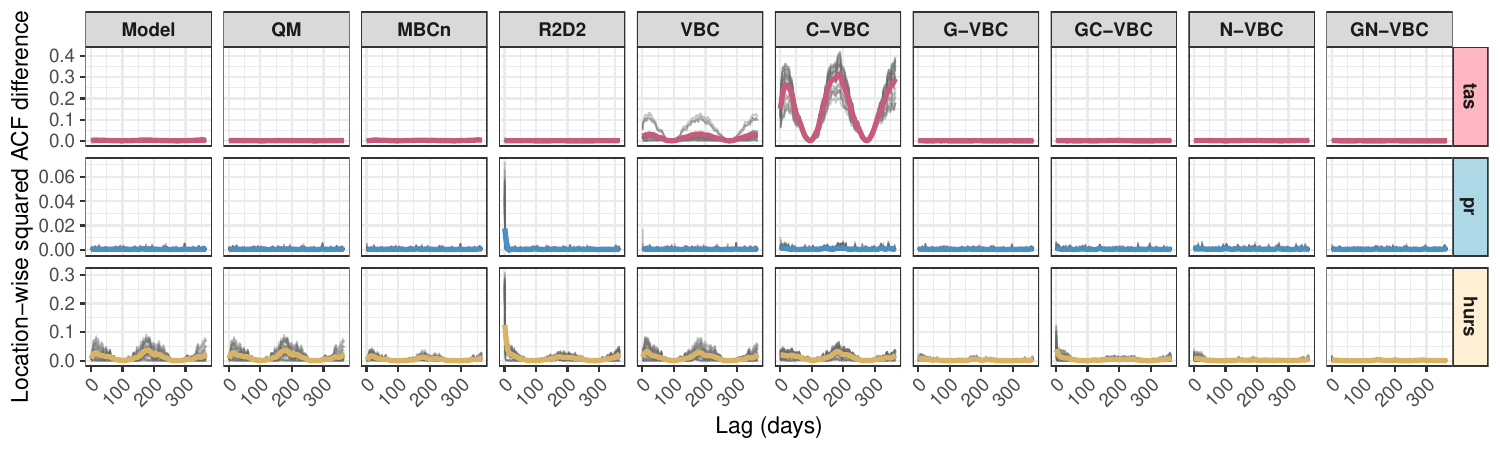}%
  }

  \medskip

  \subfloat[\label{fig:acf_30}]{%
    \includegraphics[width=\textwidth]
      {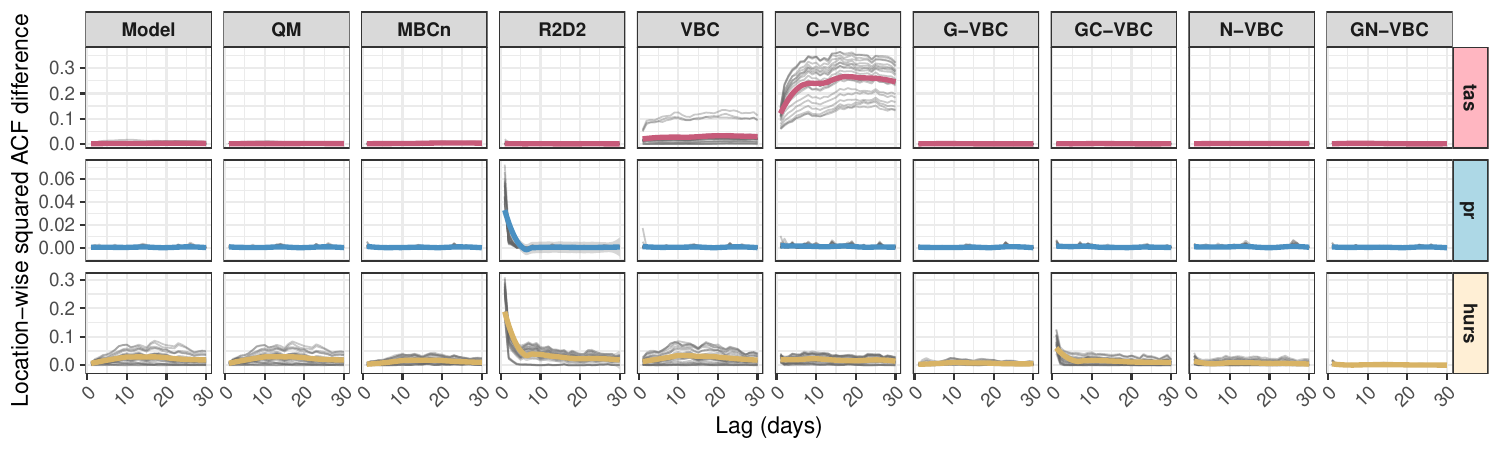}%
  }

  \caption{Lag- and location-wise squared differences in ACF values with
  respect to the reference projections, shown by variable and method for
  365 days (Panel~a) and 30 days (Panel~b). Grey lines show the squared
  differences for individual locations; smoothed lines show the mean over
  locations.}
  \label{fig:acf}
\end{figure*}

To quantify temporal consistency, we evaluate the autocorrelation function (ACF) at lags $t = 1,\dots,365$ for each variable-location pair. For each lag, we compute the squared difference between the ACF of the bias-corrected series and that of the reference projection, yielding a lag-specific measure of temporal distortion; see \cref{fig:acf_365} for lags up to one year and \cref{fig:acf_30} for short-term dependence up to 30 days. Summaries for all variables are in the supplementary material (\cref{app:tables}).

For near-surface temperature, the raw climate model output already reproduces the temporal autocorrelation structure of the reference reasonably well, leaving little scope for improvement, and QM, MBCn, and R2D2 all show similar performance. VBC applied independently at each location shows some degradation of temporal correlations at individual sites. Applying VBC jointly across locations without accounting for seasonality (C-VBC) heavily distorts the ACF, with pronounced peaks at seasonal lags, around two weeks, six months, and one year. Both the GAM decomposition and NVCs effectively remove these artifacts; G-VBC, GC-VBC, N-VBC and GN-VBC closely align with the reference ACF across all lags.

For precipitation, temporal differences are generally smaller across all methods. The uncorrected model output already shows low ACF deviations, and both QM and MBCn preserve this behavior. Although R2D2 preserves inter-variable and spatial correlations well, it clearly struggles to maintain temporal consistency, especially short-term dependencies (\cref{fig:acf_30}). The performance of all VBC-based methods is similar, with GN-VBC showing the lowest variability.
%Similarly, VBC-based approaches applied to all locations combined without NVCs (C-VBC and GC-VBC) show medium discrepancies at very short lags, suggesting slight distortion of short-term temporal consistency. Incorporating NVCs removes these early-lag discrepancies (GN-VBC).

Relative humidity presents the most challenging case. The raw model output clearly deviates from the reference ACF, exhibiting a seasonal pattern similar to that observed for temperature under joint VBC corrections (C-VBC), though less pronounced. Similar patterns are shown by QM and location-based VBC, while MBCn partially improves them without fully eliminating seasonal artifacts. Again, R2D2 distorts short-term temporal dependencies and exhibits similar seasonal patterns as MBCn. The same applies to VBC-based methods that ignore seasonality. In contrast, all methods incorporating the GAM decomposition and/or NVCs yield substantial improvements. Among these, the proposed GN-VBC approach achieves the closest agreement with the reference ACF, indicating again a synergistic effect between seasonal adjustment and structured multivariate dependence modelling.
 

%% file: sections/06_discussion.tex
% Important observations:
% \begin{itemize}
%     \item DONE Discuss other direction of the merge ("var-loc")
%     \item DONE Discuss choice of bridging location
%     \item Discuss difference between RCP8.5 and actual observed climate $\Rightarrow$ huge difference, that's why our delta approach does not improve the fit.
%     \item DONE CUVEE may be used in other domains
% \end{itemize}

% Limitations of our approach:
% \begin{itemize}
%     \item DONE Current implementation only suitable for gridded data (but possible kNN approach for valid connections between stations?)
%     \item DONE GAMs model the mean $\rightarrow$ What about extremes?
% \end{itemize}

\section{Discussion}\label{sec:dis}

%Statistical bias correction of climate model output increasingly relies on multivariate approaches to preserve dependence structures. In this work, we proposed a framework combining GAM-based spatiotemporal decomposition with Nested Vine Copulas (NVC) to model dependence across variables and locations. In an application to daily observations of five climate variables over the canton of Vaud, Switzerland, the method improves the preservation of inter-variable, spatial, and temporal dependencies relative to empirical quantile mapping and established multivariate approaches such as MBCn \citep{cannon2018multivariate}, R2D2 \citep{vrac2018multivariate}, and VBC \citep{funk2025}. An additional advantage of the approach is the interpretable graphical representation of dependencies between variable–location pairs.
With the aim of advancing operational climate projections for Switzerland (an inherently challenging task given the country’s complex topography) we propose a new multivariate bias correction method that captures inter-variable and spatiotemporal dependence more explicitly and flexibly than existing approaches, and demonstrate its strong performance on our case study. The approach builds on generalised additive models (GAMs) and newly developed nested vine copulas (NVCs). In the current implementation, location-specific vine copulas are merged through a variable-based bridging vine defined at a selected spatial location. Preliminary experiments with the inverse merging strategy, i.e.,~constructing per-location vines across variables and merging them through a spatial bridging vine, performed less favorably. This likely reflects the fact that individual climate variables typically exhibit stronger spatial dependence than multiple variables observed at a single location. A systematic comparison of alternative merging strategies remains a promising avenue for future research.

Several extensions of the framework are possible. The present implementation is tailored to gridded climate data and would require additional development to accommodate irregularly spaced locations \citep[e.g.,][]{worku2019statistical}. In addition, the bridging location is currently chosen heuristically; evaluating multiple candidates and selecting the optimal one according to objective criteria may further improve performance. However, we expect the choice of bridging location to have little effect on performance when inter-variable dependencies are similar across locations.

The GAM decomposition used in this study models smooth spatiotemporal variations in the mean but does not explicitly target tail behavior. Understanding how the proposed method affects extremes, particularly under nonstationary dependence structures, therefore remains an important direction for future work. Incorporating additional covariates, such as altitude, into the GAM formulation may also improve the representation of spatial variability.

Although motivated by climate bias correction, the NVC framework is broadly applicable to multivariate data with hierarchical dependence structures. Potential applications include joint modelling of outcomes across multiple hospitals in clinical studies or financial risk assessment across countries \citep[e.g.,][]{allen2017risk}. Beyond dependence estimation, NVCs combined with suitable marginal models also allow synthetic data generation, suggesting potential applications in stochastic weather generation \citep{claassen2025synthetic}. Overall, the results illustrate the potential of hierarchical vine constructions as a flexible and interpretable approach for modelling complex multivariate dependence structures.
% In summary, the proposed framework shows that jointly addressing spatiotemporal structure and multivariate dependence is key to improving bias correction of climate projections. The integration of GAM-based decomposition with hierarchical vine copula merging offers a flexible and interpretable approach that substantially advances current multivariate bias-correction methodologies.

%% file: sections/Appendix.tex
\section{Mathematical formulation of NVC}\label{app:math}
Remember that, in the undirected context, a spanning tree is a connected and acyclic graph.
For a given level $j \geq 1$, we denote by $E_j^{(b)}$ an arbitrary set of bridging edges connecting nodes in $N_j^{(1)} \cup N_j^{(2)} \cup E_{j-1}^{(b)}$, with $E_0^{(b)} = \emptyset$.
\begin{assumption}[Tree property]\label{assumption:tree}
  For every $j\geq 1$, $E_j^{(b)}$ satisfies the tree property if the graph $\left (N_j^{(1)} \cup N_j^{(2)} \cup  E_{j-1}^{(b)}, E_j^{(1)} \cup E_j^{(2)} \cup E_j^{(b)}\right )$, i.e.,~adding $E_j^{(b)}$ to $E_j^{(1)} \cup E_j^{(2)}$, is a spanning tree.
\end{assumption}

\begin{assumption}[Proximity condition]\label{assumption:proximity}
  For every $j \geq 2$, $E_j^{(b)}$ satisfies the proximity condition if each bridging edge $e \in E_j^{(b)}$ connects two nodes $n_1, n_2\in N_j =  E_{j-1}^{(1)} \cup E_{j-1}^{(2)} \cup E_{j-1}^{(b)}$ such that $|n_1 \cap n_2| = 1$.
\end{assumption}

\begin{theorem}\label{theorem:rvine}
  If $E_j^{(b)}$ satisfies \cref{assumption:tree} and \cref{assumption:proximity} for every $j \geq 1$, then
  the merged sequence $\mathcal{V} = (\mathcal{T}_1, \dots, \mathcal{T}_{d_1+d_2-1})$ with node and edge sets
  \begin{align*}
    N_j = N_j^{(1)} \cup N_j^{(2)} \cup E_{j-1}^{(b)}, \quad E_j = E_j^{(1)} \cup E_j^{(2)} \cup E_{j}^{(b)},
  \end{align*}
  is an R-vine tree sequence on $d_1 + d_2$ elements.
\end{theorem}

The proof is straightforward using induction.
By \cref{assumption:tree}, $\mathcal{T}_1$ is a connected and acyclic graph and therefore a valid first tree of an R-vine tree sequence on $d_1+d_2$ nodes. Assume that for some $2 \leq j \leq d_1 + d_2 -1$ the merged tree $\mathcal{T}_j$ is a valid tree with node set $N_{j+1} = E_{j} = E_{j}^{(1)} \cup E_j^{(2)} \cup E_j^{(b)}$. By \cref{assumption:proximity} every bridging edge in $E_{j+1}^{(b)}$ connects two nodes in $N_{j+1}$ that share exactly one variable, and the edges inherited from $E_{j+1}^{(1)}$ and $E_{j+1}^{(2)}$ already satisfy the proximity condition. \cref{assumption:tree} applied to level $j+1$ guarantees that the graph with edges $E_{j+1}$ is connected and acyclic, hence a valid tree. This completes the induction.

\begin{lemma}\label{lemma:graph}
  For each level $j \geq 1$ we construct a candidate tree $\mathcal{T}_j$ as follows:
  \begin{enumerate}
    \item For $j = 1$ the node set is defined as $N_1 = N_1^{(1)} \cup N_1^{(2)}$. The bridging edge $E_1^{(b)} = \{e_1^{(b)}\}$ is an element of the candidate set $C_1$ defined as
      \begin{align*}
        C_1 = \left \{\{a,b\} \subset N_1: a \in N_1^{(1)}, b \in N_1^{(2)} \right \}.
      \end{align*}
    \item Iteratively, for $j \geq 2$, set $N_j = N_j^{(1)} \cup N_j^{(2)} \cup E_{j-1}^{(b)}$ and define the candidate set $C_j$ that ensures the proximity condition as
      \begin{align*}
        C_j = \left \{ \right \{a,b\} \in N_j: a \in E_{j-1}^{(b)}, |a \cap b| = 1\}.
      \end{align*}
      Since the graph $\left (N_j,  E_j^{(1)} \cup E_j^{(2)} \cup C_j \right )$ is connected by construction, there exists a subset $E_j^{(b)}$ such that a graph with nodes $N_j$ and edges $E_j = E_j^{(1)} \cup E_j^{(2)} \cup E_j^{(b)}$ is acyclic, i.e.,~a valid tree.
  \end{enumerate}
\end{lemma}

The lemma follows directly from a well-known result in graph theory that states that every connected graph contains a spanning tree \citep{diestel2025graph}.
$\mathcal{T}_1$ is spanning tree by construction. Note that this construction is not unique and is highly dependent on the choice of bridging edges from the set of candidates. For instance, in our framework, the bridging edges can be chosen manually, randomly, or optimizing a performance measure.

\subsection{Implementation algorithm for NVC}
\label{app:nvc-algorithm}

This appendix describes the merging routine in algorithmic terms, implemented in C++ and accessible in the R package \texttt{GNVBC} available on GitHub \pkgcite.
In the implementation, an edge in tree level $t$ is represented as
\[
  (a,b\mid C), \qquad a,b \in \{1,\dots,d\},\; C\subseteq\{1,\dots,d\},\; a\notin C,\; b\notin C,
\]
where $a,b$ are conditioned variables and $C$ is the conditioning set.  For $t=1$, $C=\varnothing$.
For any edge $e=(a,b\mid C)$, define its variable support $V(e)=\{a,b\}\cup C$.
As discussed, at level $t\ge 2$, nodes are previous-level edges, i.e., each node corresponds to one edge in level $t-1$.
Let $\mathcal V^{(1)},\dots,\mathcal V^{(m)}$ be a list of $m\ge 2$ valid vine structures, possibly with different dimensions/truncation levels, with $T^{(i)}$ the truncation level of the structure $\mathcal V^{(i)}$.
The truncation level of the merged vine $\mathcal V$ is $T=\max_i T^{(i)}$, namely the largest truncation level among the inputs, and its dimension $d$ is given from the union of variable indices appearing in level-1 edges across all input vines, that is
\[
  d=\left|\bigcup_{e\in E_1^{(1)}\cup\cdots\cup E_1^{(m)}}V(e)\right|.
\]
Beyond the truncation level, all pair-copulas are set to the independence copula, as in standard truncated vine copulas. For each level $t=1,\dots,T$, $\mathcal V$ is initialised by setting
\[
  E_t \leftarrow \bigcup_{i: t\le T^{(i)}} E_t^{(i)},
\]
implemented as concatenation of edge lists.
This preserves all within-vine edges exactly and does not yet enforce that each level is a spanning tree of size $d-t$.
The next and most important phase of the implementation is the completion by bridging edges.
For each level $t=1,\dots,T$, define target edge count
\[
  |E_t|_{\text{target}} = d-t.
\]
If current $|E_t|$ already matches target, i.e., if the merged input already has the required number of edges, no action is needed. Otherwise, level completion is performed by Algorithm~\ref{alg:fill-missing-driver}: for each level $t$, it computes the target size $m_t=d-t$ and calls three subroutines:
\begin{itemize}
  \item Algorithm~\ref{alg:fill-missing-reconstruct}: builds the level-specific node representation and initialises the disjoint-set union (DSU; \citealp{galler1964,tarjan1975}) state from already present edges.
  \item Algorithm~\ref{alg:fill-missing-candidates}: enumerates admissible bridging edges by checking the proximity condition (\cref{assumption:proximity}), implemented as $|V(u)\cap V(v)|=|V(u)|-1$.
  \item Algorithm~\ref{alg:fill-missing-greedy}: adds candidates greedily while enforcing the tree property (\cref{assumption:tree}) via DSU cycle checks, stopping when a spanning tree is reached or no admissible candidate remains.
\end{itemize}
At each processed level, the routine returns an acyclic connected graph on $N_t$ with exactly $|N_t|-1=d-t$ edges, i.e., a spanning tree consistent with the implemented proximity rule.
Note that the initial bridging choice in level 1 can be user-specified, random, or criterion-driven before running completion.
The implemented completion is then deterministic given candidate ordering; replacing the ordering yields alternative valid merged vines.
The rest of this subsection is as follows. In \cref{sec:complexity-analysis}, we give per-level and overall complexity bounds. In \cref{sec:fill-missing-reconstruct,sec:fill-missing-candidates,sec:fill-missing-greedy}, we detail the three subroutines called by Algorithm~\ref{alg:fill-missing-driver}.

\begin{algorithm}[t]
  \caption{Level-wise completion by bridging edges}
  \label{alg:fill-missing-driver}
  \textbf{Input:} partially merged edge lists $(E_t)_{t=1}^T$, global dimension $d$.\\
  \textbf{Output:} completed edge lists $(E_t)_{t=1}^T$ such that $|E_t|=d-t$ when feasible.
  \begin{algorithmic}[1]
    \For{$t=1$ \textbf{to} $T$}
    \State $m_t \gets d-t$
    \If{$|E_t|=m_t$} \State \textbf{continue} \EndIf
    \State $(N_t,\text{DSU}) \gets \Call{ReconstructAndInitDSU}{t, E_{t-1}, E_t}$ \Comment{Algorithm~\ref{alg:fill-missing-reconstruct}}
    \State $C_t \gets \Call{BuildCandidates}{N_t, E_t}$ \Comment{Algorithm~\ref{alg:fill-missing-candidates}}
    \State $E_t \gets \Call{GreedyAugment}{E_t,C_t,\text{DSU},m_t}$ \Comment{Algorithm~\ref{alg:fill-missing-greedy}}
    \If{$|E_t|<m_t$}
    \State \textbf{fail} at level $t$
    \EndIf
    \EndFor
    \State \textbf{return} $(E_t)_{t=1}^T$
  \end{algorithmic}
\end{algorithm}

\subsubsection{Complexity Analysis}\label{sec:complexity-analysis}
Let $n_t$ be the number of nodes at level $t$; the target number of edges is therefore $n_t-1$.
\begin{itemize}
  \item In reconstruction/linking between levels (Algorithm~\ref{alg:fill-missing-reconstruct}), each edge from level $t$ is indexed by two keys,
    $(a, C\cup\{b\})$ and $(b, C\cup\{a\})$, in an associative map \citep{cormen2022}. This supports incident-node lookup for an edge $(a,b\mid C)$ in $O(\log |E_t|)$ time; overall map build + lookups contribute $O(n_t\log n_t)$ per level.
  \item Candidate generation  scans $O(n_t^2)$ node pairs; each pair uses set intersection/difference on supports of size $t$, giving $O(t)$ using an appropriate set representation.
  \item DSU operations (Algorithms~\ref{alg:fill-missing-reconstruct} and~\ref{alg:fill-missing-greedy}) are near-constant amortised, $O(\alpha(n_t))$ \citep{tarjan1975}, so cycle checks are effectively linear in the number of attempted insertions.
\end{itemize}
Hence a compact per-level bound is $O\!\left(n_t^2\, t\right)$
with candidate generation as the dominant term.
Summing across levels gives the overall bound $O\!\left(\sum_{t=1}^{T} n_t^2\, t\right)$.
For non-truncated vines, namely when filling is performed until level $d-1$, this yields $O(d^4)$.
Currently, the algorithm proceeds at each step in Kruskal-style like \citep{kruskal1956} by enumerating the all edges. A promising improvement is to replace the quadratic pair scan by a component-centric, bucketed Boruvka-style completion \citep{nevsetvril2001otakar}, in which nodes are indexed by shared $(t-1)$-subsets and each connected component selects one outgoing admissible edge per round; if this can be implemented without hidden quadratic overhead, the overall complexity may plausibly be reduced to $O(d^3\log d)$.

\subsubsection{Node Reconstruction and DSU initialisation}\label{sec:fill-missing-reconstruct}
Algorithm~\ref{alg:fill-missing-reconstruct} is the bridge between vine semantics and graph operations.
For $t=1$, nodes are variables; for $t\ge2$, nodes are edges from the previous level, represented through supports $V(e)$.
The DSU is seeded with existing edges so that any cycle inherited from the current partial graph is detected immediately.

\begin{algorithm}[t]
  \caption{Node reconstruction and DSU initialisation}
  \label{alg:fill-missing-reconstruct}
  \textbf{Input:} level index $t$, previous-level edges $E_{t-1}$ (unused if $t=1$), current edges $E_t$.\\
  \textbf{Output:} node set $N_t$ with supports $V(\cdot)$ and DSU initialised with existing edges.
  \begin{algorithmic}[1]
    \If{$t=1$}
    \State Create one node per variable index; set node support size to $1$
    \Else
    \State Initialise $N_t$ by creating one node per edge in $E_{t-1}$
    \State Assign node support as $V(e)$ for the corresponding edge $e$
    \EndIf
    \State Initialise DSU on nodes in $N_t$
    \For{each edge in $E_t$}
    \If{edge endpoints are already in same DSU component}
    \State \textbf{fail} (cycle detected in current level)
    \Else
    \State Union the two endpoint components in DSU
    \EndIf
    \EndFor
    \State \textbf{return} $(N_t,\text{DSU})$
  \end{algorithmic}
\end{algorithm}

\subsubsection{Candidate Generation from Support Overlaps}\label{sec:fill-missing-candidates}

Algorithm~\ref{alg:fill-missing-candidates} enforces the proximity condition using only set operations.
For each non-adjacent node pair, it computes the shared support and accepts the pair exactly when supports differ by one variable.
The two differing variables define $(a,b)$, while the shared part defines the conditioning set $S$.

\begin{algorithm}[t]
  \caption{Candidate bridge generation from support overlaps}
  \label{alg:fill-missing-candidates}
  \textbf{Input:} node set $N_t$ with supports $V(\cdot)$, current edge set $E_t$.\\
  \textbf{Output:} candidate list $C_t$.
  \begin{algorithmic}[1]
    \State $C_t \gets \varnothing$
    \For{each unordered non-adjacent pair $(u,v)\in N_t$}
    \State $S \gets V(u)\cap V(v)$
    \If{$|S|=|V(u)|-1$}
    \State $a \gets$ unique element of $V(u)\setminus S$
    \State $b \gets$ unique element of $V(v)\setminus S$
    \State Append candidate $(u,v; a,b\mid S)$ to $C_t$
    \EndIf
    \EndFor
    \State \textbf{return} $C_t$
  \end{algorithmic}
\end{algorithm}

\subsubsection{Greedy Augmentation under Acyclicity}\label{sec:fill-missing-greedy}

Algorithm~\ref{alg:fill-missing-greedy} is a Kruskal-style \citep{kruskal1956} completion step over the admissible candidates. The function find$(\cdot)$ follows the chain of parent pointers from a specified query node until it reaches a root element. This root element signifies the set to which the node belongs, and it may also be the node itself. Candidates are processed in implementation order (stack order), and each edge is accepted only if it connects two distinct DSU components.
Hence every accepted edge reduces the number of connected components and never creates a cycle.

\begin{algorithm}[t]
  \caption{Greedy augmentation under acyclicity}
  \label{alg:fill-missing-greedy}
  \textbf{Input:} current edges $E_t$, candidate list $C_t$, DSU, target size $m_t$.\\
  \textbf{Output:} updated $E_t$.
  \begin{algorithmic}[1]
    \While{$|E_t|<m_t$ \textbf{and} $C_t\neq\varnothing$}
    \State Pop one candidate $(u,v; a,b\mid S)$ from $C_t$
    \If{$\mathrm{find}(u)=\mathrm{find}(v)$}
    \State \textbf{continue}
    \Else
    \State Add edge $(a,b\mid S)$ to $E_t$
    \State Union endpoint components of $u,v$ in DSU
    \EndIf
    \EndWhile
    \State \textbf{return} $E_t$
  \end{algorithmic}
\end{algorithm}

\section{Simulation Studies}\label{app:sim}
We perform experiments in order to evaluate both our vine merging algorithm NVC outside of the bias correction framework, and our proposed GN-VBC for artificially created climate data. 
\subsection{Experiments on NVCs}\label{app:sim:nvc}

To evaluate the benefits of fitting NVCs rather than standard vine
copulas, we simulate hierarchically structured data in the
variable-location format used in climate projections. Although
motivated by this application, the simulation framework can also
represent other data with a comparable hierarchical structure.

We consider data with uniform margins for different numbers of variables ($\{2,3,4,5\}$) and locations ($\{4,9,16,25\}$), with locations arranged on square grids of sizes $2\times2$, $3\times3$, $4\times4$, and $5\times5$, respectively. We first construct an NVC structure by drawing a random spatial vine structure for each variable, a random inter-variable vine structure that determines the bridging edges, and a random bridging location. For each edge of the resulting structure, we then randomly select a pair-copula family from the Gaussian, Clayton, Frank, Gumbel, and Joe families, a rotation where applicable, and a value of Kendall's $\tau$. When choosing Kendall's $\tau$, we distinguish between within-and between-variable dependencies to represent a hierarchical dependence structure motivated by climate projections.

From the resulting vine copula, we generate training samples of sizes $n\in\{500,2000,5000\}$ and fit the following three models:
\begin{itemize}
    \item \textbf{Oracle vine copula:}
    The true vine copula structure is supplied, and only the corresponding pair-copulas are estimated. This model serves as a benchmark with known structure.

    \item \textbf{NVC:}
    The model is fitted using our NVC algorithm, described in \cref{sec:nvc,app:math}, with the numbers of variables and locations and the bridging location supplied.

    \item \textbf{Standard vine copula:}
    A vine copula is fitted jointly to all variable-location pairs without imposing the nested structure.
\end{itemize}

We consider both non-parametric and parametric pair-copula estimation. In the application in \cref{sec:application}, however, all vine copulas are estimated non-parametrically.

For each configuration, we perform 100 simulation replicates and evaluate the fitted models using a Monte Carlo approximation of the Kullback-Leibler (KL) divergence based on $10,000$ observations. \Cref{tab:kl-comparison} reports the median and quartiles of the resulting KL values. Across the configurations considered, parametric fitting yields lower median KL divergences than non-parametric fitting. This is consistent with the construction of the data-generating copulas from parametric pair-copula families. For both fitting approaches, the median KL divergences decrease with the training sample size and generally increase with the number of variables and locations.

\begin{figure}
    \centering
    \includegraphics[width=\linewidth]{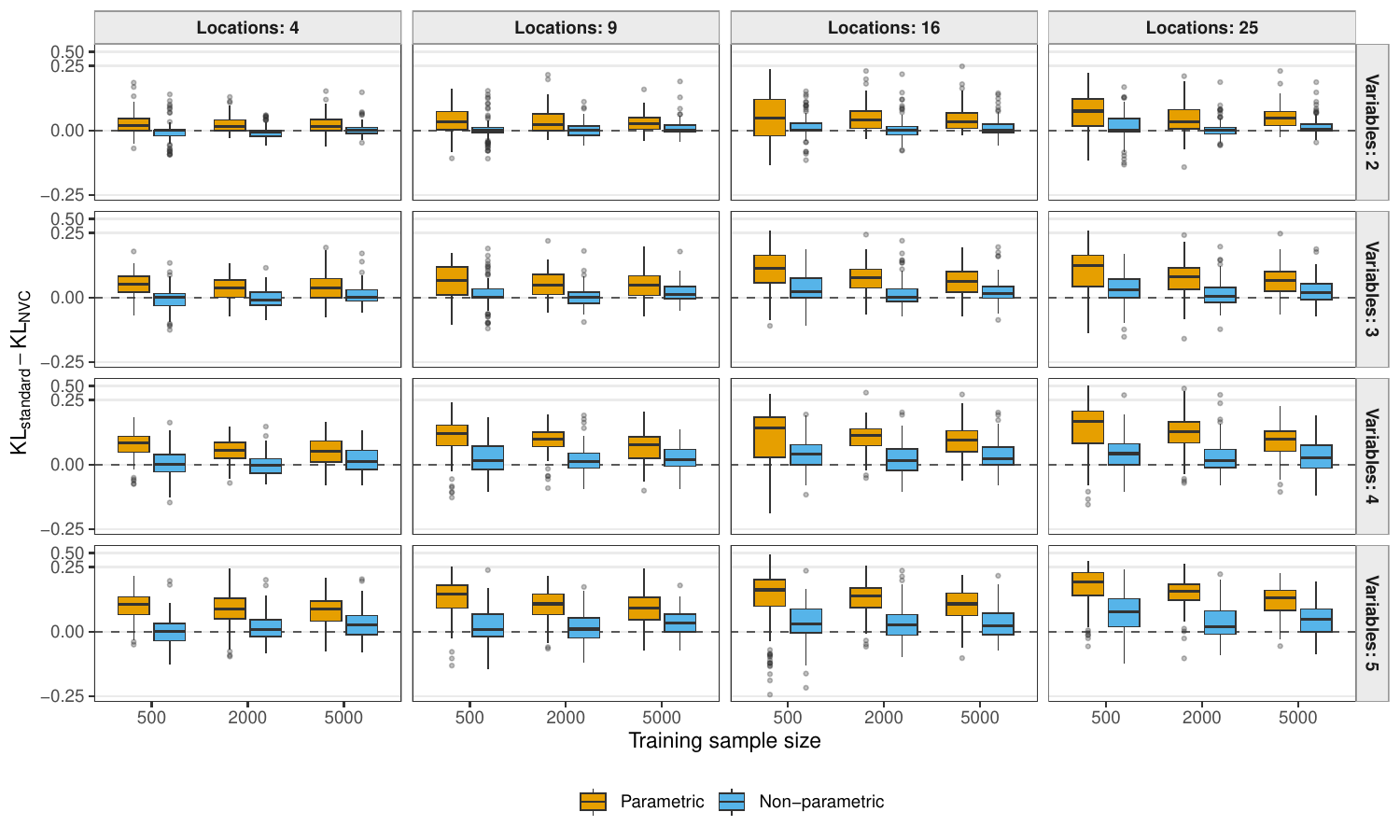}
    \caption{Paired differences $KL_{\mathrm{standard}}-KL_{\mathrm{NVC}}$ across 100 simulation replicates for three training sample sizes and two pair-copula estimation approaches. Columns indicate the number of locations, and rows indicate the number of variables. Positive values favour the estimated NVC, while negative values favour the standard vine.}
    \label{fig:sim_study_nvc}
\end{figure}

To examine the effect of imposing the nested structure, \cref{fig:sim_study_nvc} displays the paired differences $KL_{\mathrm{standard}}-KL_{\mathrm{NVC}}$, calculated within each replicate. Under parametric fitting, the median difference is positive in every configuration and generally increases with the number of variables and locations. The advantage is often most pronounced for the smallest training sample size, particularly in larger dimensions. Under non-parametric fitting, the differences are generally smaller. In lower-dimensional configurations, the median differences are close to zero, indicating similar performance of the two approaches. The benefits of imposing the nested structure become more apparent as the number of variables and locations increases, although the magnitude of the improvement varies across replicates and sample sizes. In the largest configurations, these benefits are particularly visible for $n=500$.

Our application in \cref{sec:application} involves five variables
at 22 locations. The closest simulated configuration in terms
of these counts is therefore the setting with five variables
and 25 locations, shown in the lower-right panel of
\cref{fig:sim_study_nvc}. This configuration exhibits positive
median paired differences for both fitting approaches and all
three training sample sizes.

\begin{table}[p]

\scriptsize
\setlength{\tabcolsep}{2pt}
\renewcommand{\arraystretch}{1.0}

% \begin{adjustbox}{
%   max totalsize={\textwidth}{0.82\textheight},
%   center
% }
\begin{tabular}{@{}rrrlll@{\hspace{5pt}}lll@{}}

\toprule
& & & \multicolumn{3}{c}{Non-parametric}
    & \multicolumn{3}{c}{Parametric} \\
\cmidrule(lr){4-6}\cmidrule(lr){7-9}
Vars. & Locs. & $n$
& Oracle & Standard & NVC
& Oracle & Standard & NVC \\
\midrule

2 & 4 & 500
& 0.72 [0.61, 0.82] & 0.74 [0.62, 0.85] & 0.74 [0.63, 0.85]
& 0.03 [0.02, 0.04] & 0.07 [0.04, 0.12] & 0.06 [0.03, 0.11] \\
2 & 4 & 2000
& 0.47 [0.39, 0.59] & 0.49 [0.40, 0.60] & 0.49 [0.40, 0.60]
& 0.01 [0.01, 0.01] & 0.03 [0.02, 0.08] & 0.02 [0.01, 0.07] \\
2 & 4 & 5000
& 0.37 [0.30, 0.49] & 0.39 [0.33, 0.49] & 0.39 [0.32, 0.49]
& 0.00 [0.00, 0.00] & 0.02 [0.01, 0.06] & 0.01 [0.00, 0.06] \\
\addlinespace[2pt]

2 & 9 & 500
& 2.14 [1.96, 2.31] & 2.24 [2.05, 2.49] & 2.24 [2.04, 2.49]
& 0.21 [0.20, 0.24] & 0.44 [0.32, 0.59] & 0.44 [0.31, 0.60] \\
2 & 9 & 2000
& 1.40 [1.25, 1.55] & 1.50 [1.33, 1.66] & 1.50 [1.33, 1.65]
& 0.05 [0.05, 0.06] & 0.24 [0.14, 0.36] & 0.23 [0.14, 0.37] \\
2 & 9 & 5000
& 1.13 [0.97, 1.27] & 1.22 [1.05, 1.37] & 1.22 [1.05, 1.37]
& 0.02 [0.02, 0.02] & 0.20 [0.09, 0.32] & 0.19 [0.08, 0.31] \\
\addlinespace[2pt]

2 & 16 & 500
& 4.08 [3.79, 4.32] & 4.44 [4.16, 4.72] & 4.43 [4.16, 4.72]
& 0.58 [0.55, 0.63] & 1.10 [0.95, 1.33] & 1.08 [0.93, 1.34] \\
2 & 16 & 2000
& 2.69 [2.53, 2.86] & 2.98 [2.85, 3.13] & 2.97 [2.85, 3.13]
& 0.15 [0.14, 0.16] & 0.64 [0.45, 0.78] & 0.62 [0.43, 0.76] \\
2 & 16 & 5000
& 2.16 [1.99, 2.33] & 2.44 [2.30, 2.58] & 2.44 [2.28, 2.59]
& 0.06 [0.06, 0.07] & 0.55 [0.39, 0.68] & 0.52 [0.38, 0.67] \\
\addlinespace[2pt]

2 & 25 & 500
& 6.61 [6.33, 6.92] & 7.40 [6.99, 7.67] & 7.38 [6.99, 7.68]
& 1.02 [0.97, 1.07] & 2.14 [1.93, 2.32] & 2.12 [1.92, 2.31] \\
2 & 25 & 2000
& 4.37 [4.18, 4.62] & 5.01 [4.73, 5.24] & 5.01 [4.73, 5.22]
& 0.26 [0.25, 0.28] & 1.18 [0.97, 1.45] & 1.17 [0.95, 1.46] \\
2 & 25 & 5000
& 3.48 [3.31, 3.73] & 4.11 [3.84, 4.33] & 4.10 [3.84, 4.34]
& 0.11 [0.10, 0.11] & 1.02 [0.86, 1.23] & 1.01 [0.84, 1.22] \\

\midrule

3 & 4 & 500
& 1.12 [0.94, 1.26] & 1.15 [1.00, 1.28] & 1.15 [1.01, 1.29]
& 0.06 [0.05, 0.07] & 0.13 [0.09, 0.21] & 0.11 [0.08, 0.20] \\
3 & 4 & 2000
& 0.74 [0.61, 0.88] & 0.77 [0.64, 0.89] & 0.77 [0.65, 0.89]
& 0.01 [0.01, 0.02] & 0.06 [0.04, 0.13] & 0.06 [0.03, 0.11] \\
3 & 4 & 5000
& 0.60 [0.49, 0.73] & 0.64 [0.52, 0.75] & 0.63 [0.52, 0.75]
& 0.00 [0.00, 0.01] & 0.06 [0.03, 0.13] & 0.05 [0.02, 0.11] \\
\addlinespace[2pt]

3 & 9 & 500
& 3.18 [2.90, 3.42] & 3.39 [3.16, 3.60] & 3.39 [3.15, 3.60]
& 0.35 [0.32, 0.38] & 0.72 [0.56, 0.92] & 0.68 [0.53, 0.90] \\
3 & 9 & 2000
& 2.10 [1.87, 2.29] & 2.28 [2.06, 2.46] & 2.28 [2.04, 2.44]
& 0.09 [0.09, 0.10] & 0.43 [0.30, 0.61] & 0.42 [0.28, 0.60] \\
3 & 9 & 5000
& 1.68 [1.48, 1.88] & 1.85 [1.69, 2.02] & 1.84 [1.68, 2.02]
& 0.04 [0.03, 0.04] & 0.36 [0.22, 0.56] & 0.35 [0.20, 0.54] \\
\addlinespace[2pt]

3 & 16 & 500
& 6.07 [5.71, 6.40] & 6.56 [6.21, 6.99] & 6.56 [6.19, 6.97]
& 0.94 [0.90, 0.99] & 1.82 [1.56, 1.99] & 1.74 [1.52, 1.94] \\
3 & 16 & 2000
& 3.96 [3.76, 4.23] & 4.38 [4.18, 4.74] & 4.36 [4.17, 4.74]
& 0.24 [0.23, 0.26] & 0.97 [0.77, 1.17] & 0.95 [0.75, 1.14] \\
3 & 16 & 5000
& 3.19 [3.02, 3.38] & 3.52 [3.34, 3.89] & 3.51 [3.32, 3.87]
& 0.10 [0.09, 0.10] & 0.79 [0.56, 0.97] & 0.76 [0.53, 0.96] \\
\addlinespace[2pt]

3 & 25 & 500
& 10.09 [9.71, 10.48] & 11.17 [10.73, 11.48] & 11.16 [10.70, 11.48]
& 1.59 [1.52, 1.65] & 3.30 [2.95, 3.55] & 3.22 [2.91, 3.48] \\
3 & 25 & 2000
& 6.68 [6.36, 7.00] & 7.54 [7.20, 7.88] & 7.53 [7.19, 7.86]
& 0.42 [0.40, 0.44] & 1.86 [1.61, 2.15] & 1.85 [1.56, 2.13] \\
3 & 25 & 5000
& 5.36 [5.08, 5.58] & 6.15 [5.89, 6.42] & 6.15 [5.86, 6.42]
& 0.17 [0.16, 0.18] & 1.57 [1.29, 1.86] & 1.58 [1.29, 1.84] \\

\midrule

4 & 4 & 500
& 1.41 [1.24, 1.61] & 1.44 [1.29, 1.66] & 1.44 [1.30, 1.66]
& 0.09 [0.07, 0.10] & 0.21 [0.16, 0.31] & 0.18 [0.13, 0.28] \\
4 & 4 & 2000
& 0.92 [0.77, 1.07] & 0.97 [0.81, 1.10] & 0.96 [0.81, 1.10]
& 0.02 [0.02, 0.03] & 0.13 [0.08, 0.21] & 0.11 [0.06, 0.20] \\
4 & 4 & 5000
& 0.73 [0.61, 0.87] & 0.79 [0.65, 0.91] & 0.78 [0.65, 0.91]
& 0.01 [0.01, 0.01] & 0.13 [0.08, 0.19] & 0.10 [0.05, 0.19] \\
\addlinespace[2pt]

4 & 9 & 500
& 4.33 [4.07, 4.57] & 4.60 [4.27, 4.85] & 4.60 [4.26, 4.84]
& 0.49 [0.46, 0.53] & 0.90 [0.78, 1.13] & 0.88 [0.71, 1.04] \\
4 & 9 & 2000
& 2.83 [2.58, 3.08] & 3.05 [2.81, 3.28] & 3.06 [2.77, 3.27]
& 0.13 [0.12, 0.14] & 0.58 [0.43, 0.72] & 0.52 [0.41, 0.69] \\
4 & 9 & 5000
& 2.30 [2.07, 2.48] & 2.50 [2.24, 2.73] & 2.49 [2.22, 2.71]
& 0.05 [0.05, 0.06] & 0.46 [0.30, 0.63] & 0.45 [0.27, 0.61] \\
\addlinespace[2pt]

4 & 16 & 500
& 8.31 [7.97, 8.65] & 9.05 [8.74, 9.34] & 9.04 [8.73, 9.34]
& 1.31 [1.23, 1.37] & 2.54 [2.21, 2.92] & 2.48 [2.16, 2.87] \\
4 & 16 & 2000
& 5.51 [5.22, 5.84] & 6.09 [5.77, 6.38] & 6.08 [5.76, 6.37]
& 0.34 [0.32, 0.36] & 1.39 [1.14, 1.66] & 1.35 [1.12, 1.61] \\
4 & 16 & 5000
& 4.42 [4.18, 4.75] & 4.97 [4.70, 5.24] & 4.96 [4.64, 5.22]
& 0.14 [0.13, 0.15] & 1.12 [0.95, 1.43] & 1.10 [0.91, 1.39] \\
\addlinespace[2pt]

4 & 25 & 500
& 13.36 [12.89, 13.94] & 14.80 [14.19, 15.56] & 14.78 [14.20, 15.55]
& 2.15 [2.08, 2.22] & 4.38 [4.03, 4.71] & 4.27 [3.91, 4.62] \\
4 & 25 & 2000
& 8.86 [8.40, 9.26] & 10.19 [9.72, 10.51] & 10.18 [9.69, 10.51]
& 0.56 [0.54, 0.59] & 2.56 [2.27, 2.93] & 2.49 [2.21, 2.88] \\
4 & 25 & 5000
& 7.15 [6.68, 7.49] & 8.22 [7.83, 8.74] & 8.18 [7.83, 8.70]
& 0.23 [0.22, 0.24] & 2.07 [1.86, 2.47] & 2.04 [1.81, 2.40] \\

\midrule

5 & 4 & 500
& 1.84 [1.70, 2.02] & 1.90 [1.73, 2.11] & 1.89 [1.73, 2.11]
& 0.12 [0.10, 0.13] & 0.30 [0.20, 0.42] & 0.26 [0.17, 0.39] \\
5 & 4 & 2000
& 1.21 [1.11, 1.38] & 1.29 [1.16, 1.44] & 1.28 [1.15, 1.44]
& 0.03 [0.02, 0.03] & 0.18 [0.12, 0.28] & 0.15 [0.08, 0.25] \\
5 & 4 & 5000
& 0.97 [0.88, 1.11] & 1.06 [0.94, 1.21] & 1.06 [0.92, 1.20]
& 0.01 [0.01, 0.01] & 0.16 [0.09, 0.25] & 0.13 [0.06, 0.23] \\
\addlinespace[2pt]

5 & 9 & 500
& 5.29 [4.95, 5.56] & 5.70 [5.35, 5.96] & 5.67 [5.34, 5.96]
& 0.64 [0.59, 0.68] & 1.33 [1.11, 1.57] & 1.26 [1.04, 1.52] \\
5 & 9 & 2000
& 3.52 [3.26, 3.71] & 3.75 [3.54, 4.00] & 3.75 [3.52, 3.98]
& 0.16 [0.15, 0.17] & 0.74 [0.56, 0.90] & 0.71 [0.53, 0.84] \\
5 & 9 & 5000
& 2.81 [2.60, 2.98] & 3.03 [2.87, 3.22] & 3.00 [2.87, 3.21]
& 0.07 [0.06, 0.07] & 0.62 [0.47, 0.80] & 0.58 [0.43, 0.77] \\
\addlinespace[2pt]

5 & 16 & 500
& 10.32 [9.91, 10.84] & 11.16 [10.76, 11.75] & 11.17 [10.75, 11.70]
& 1.68 [1.62, 1.77] & 3.16 [2.86, 3.55] & 3.07 [2.80, 3.44] \\
5 & 16 & 2000
& 6.80 [6.48, 7.17] & 7.54 [7.19, 7.92] & 7.52 [7.19, 7.88]
& 0.43 [0.41, 0.45] & 1.68 [1.43, 2.11] & 1.62 [1.38, 2.03] \\
5 & 16 & 5000
& 5.42 [5.12, 5.72] & 6.16 [5.71, 6.45] & 6.15 [5.70, 6.45]
& 0.18 [0.17, 0.19] & 1.39 [1.17, 1.72] & 1.32 [1.13, 1.66] \\
\addlinespace[2pt]

5 & 25 & 500
& 16.79 [16.48, 17.42] & 18.55 [18.00, 19.38] & 18.52 [18.01, 19.37]
& 2.75 [2.67, 2.83] & 5.52 [5.02, 6.03] & 5.40 [4.92, 5.85] \\
5 & 25 & 2000
& 11.28 [10.77, 11.67] & 12.64 [12.28, 13.15] & 12.64 [12.25, 13.17]
& 0.72 [0.70, 0.75] & 3.12 [2.81, 3.43] & 3.05 [2.74, 3.35] \\
5 & 25 & 5000
& 9.03 [8.61, 9.38] & 10.28 [10.02, 10.74] & 10.25 [10.01, 10.73]
& 0.29 [0.28, 0.30] & 2.52 [2.22, 2.87] & 2.50 [2.14, 2.80] \\

\bottomrule
\end{tabular}
% \end{adjustbox}
\centering
\caption{KL divergence for non-parametric and parametric fitting
across simulation configurations. Entries report the median
[$Q_1$, $Q_3$] across replicates, rounded to two decimal places.
Vars.\ and Locs.\ denote the numbers of variables and locations,
respectively, and $n$ is the training sample size.
Values displayed as 0.00 may be positive before rounding.}
\label{tab:kl-comparison}
\end{table}

\subsection{Experiments on GN-VBC}\label{app:sim_study_mbc}
The aim of this simulation study is to assess whether the proposed bias-correction approach preserves spatial, inter-variable, and temporal dependence structures in a controlled setting. To this end, we generate artificial multivariate time-series data representing two variables observed at two spatial locations, resulting in a four-dimensional process with one dimension per variable--location pair. Separate reference and model data sets are generated, allowing for systematic differences in marginal and cross-sectional dependence structures.

Temporal dependence is introduced through latent Gaussian processes evolving according to a first-order autoregressive model. Let $t=1,\dots,T$ with $T=200$, and let $Z_t^{(r)}, Z_t^{(m)} \in \mathbb{R}^4$ denote the latent reference and model processes at time $t$. Both processes use the same autoregressive coefficient, $\phi=0.4$:
\begin{align*}
  Z_t^{(r)} &= \phi Z_{t-1}^{(r)} + \varepsilon_t^{(r)}, \\
  Z_t^{(m)} &= \phi Z_{t-1}^{(m)} + \varepsilon_t^{(m)}.
\end{align*}
The innovation vectors are independent over time and have a cross-sectional dependence structure specified by Gaussian R-vine copulas. The vine structure represents dependence between the same variable at the two locations (cross-location dependence) and between the two variables at the same location (inter-variable dependence).

We consider two experiments. In the cross-location experiment, the model and reference cross-location Gaussian copula parameters in the pair-copula constructions in $\varepsilon_t^{(r)}$ and $\varepsilon_t^{(m)}$ vary independently over $\{0,0.1,\dots,0.9,0.99\}$, while the inter-variable parameter is fixed at $0.1$ in both data sets. In the inter-variable experiment, the model and reference inter-variable parameters vary independently over the same set, while the cross-location parameter is fixed at $0.9$ in both. The varied dependence parameters can therefore differ between the model and reference.

\begin{figure}
    \centering
    \includegraphics[width=\linewidth]{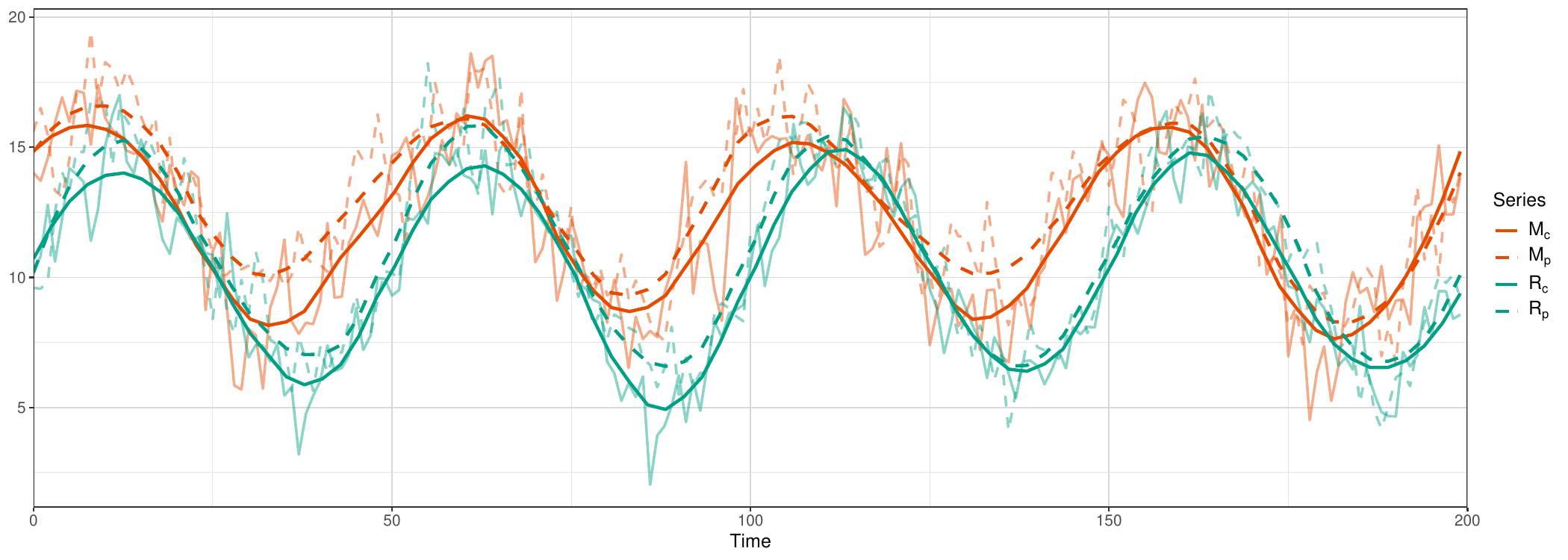}
    \caption{Example of the four simulated time series for one variable.}
    \label{fig:sim_ex}
\end{figure}

Marginal non-stationarity is introduced through time-varying means. Each variable--location series has a sinusoidal mean with four cycles over the $T=200$ time points. Let $q_t = ((t-1) \bmod 50)/50$ denote the phase within a cycle. For data set $d \in \{r,m\}$ and period $s \in \{c,p\}$, define
\[
  m_{t,k}^{(d,s)}
  =
  a_k^{(d)}
  + A_k^{(d)} \sin\!\left(2\pi(q_t+\delta_k^{(d)})\right)
  + \mathbbm{1}_{\{s=p\}}\,b_k,
\]
where $a_k^{(d)}$, $A_k^{(d)}$, and $\delta_k^{(d)}$ are the series-specific intercept, amplitude, and phase shift. These parameters, as well as the marginal scales $\sigma_k^{(d)}$, are fixed and differ between the reference and model data. The projection signal $b_k$ is shared by both. The observed series are obtained by adding the scaled latent autoregressive process to the time-varying mean:
\[
  Y_{t,k}^{(d,s)}
  =
  m_{t,k}^{(d,s)} + \sigma_k^{(d)} Z_{t,k}^{(d,s)},
  \qquad d \in \{r,m\},\quad s \in \{c,p\}.
\]
Calibration and projection series are generated from separate simulations of the corresponding autoregressive processes, each with a burn-in of 200 time points; see \cref{fig:sim_ex} for an example. The full simulation is repeated 100 times using different random seeds.

\begin{figure}
  \centering
  \includegraphics[width=\linewidth]{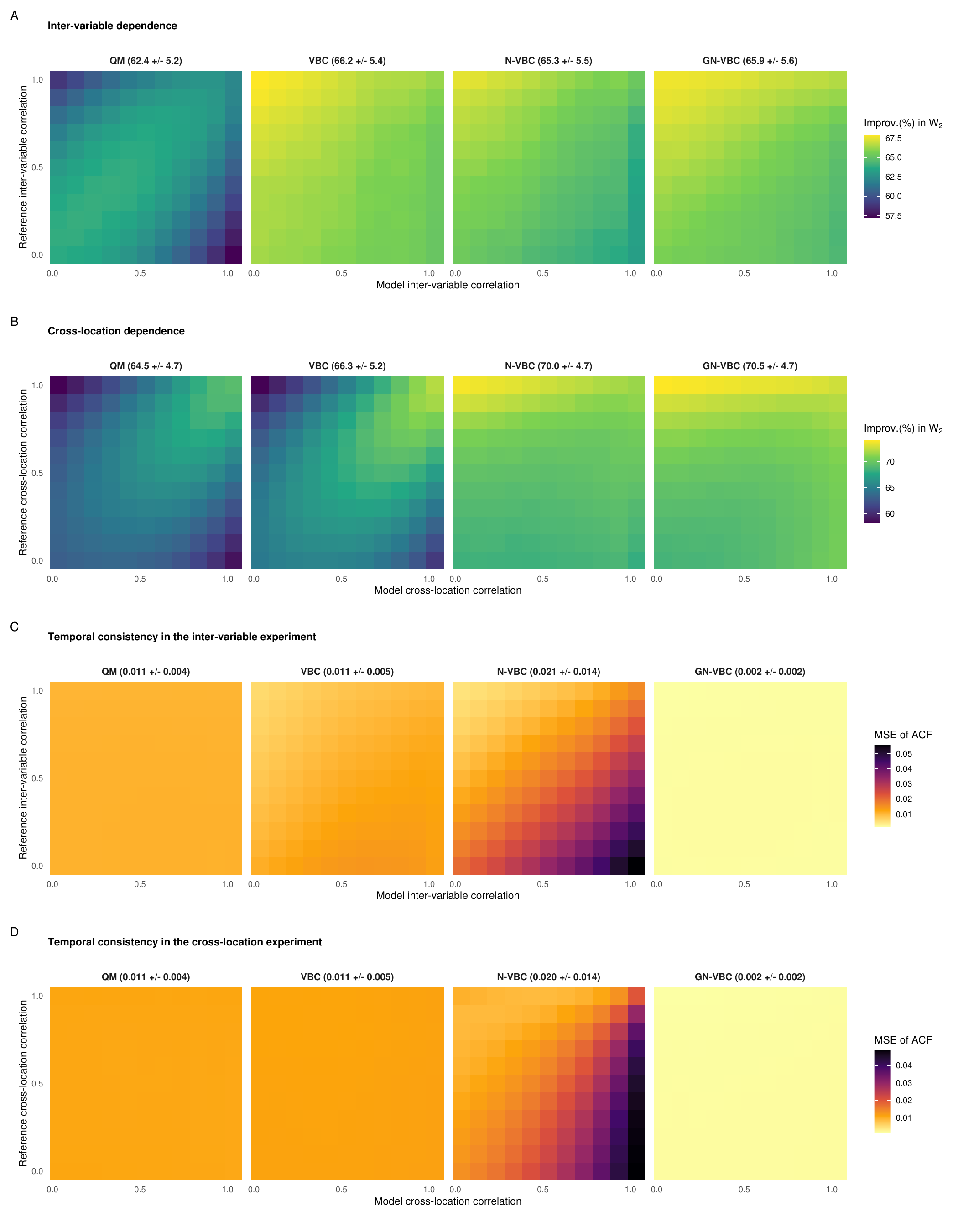}
  \caption{Simulation results for the inter-variable and cross-location experiments. Panels A and B show the percentage improvement in the four-dimensional second Wasserstein distance relative to the uncorrected model projections as the inter-variable and cross-location Gaussian copula parameters vary, respectively. Panels C and D show the MSE between the reference and corrected-series ACFs over lags 1--10 for the inter-variable and cross-location experiments, respectively. Facet titles report the mean and standard deviation across correlation scenarios.}
  \label{fig:sim_results}
\end{figure}

% \begin{figure}
%     \centering
%     \includegraphics[width=\linewidth]{images/Simulation/simulation_tau_mse.pdf}
%     \caption{Mean squared error in Kendall's $\tau$ between the reference projections and bias-corrected model projections for (A) inter-variable pairs and (B) cross-location pairs. The varied dependence parameters are Gaussian copula parameters; the other dependence parameter is held fixed as described for \cref{fig:sim_results}. Facet titles report the mean and standard deviation across correlation scenarios.}    \label{fig:sim_results_tau}
% \end{figure}
We compare four bias-correction approaches: (i) empirical QM, applied independently to each of the four time series; (ii) standard VBC, applied separately at each location using bivariate vine copulas; (iii) N-VBC, our hierarchical vine-merging extension of VBC, in which two variable-specific vines are fitted and subsequently merged into a global dependence structure; and (iv) GN-VBC, where marginal non-stationarity is first removed by fitting a GAM to each series as a function of time as introduced in \cref{sec:gam}, and NVC is then applied to the resulting PIT residuals.

Inter-variable and cross-location consistency are assessed using the multivariate second Wasserstein distance \citep{villani2008optimal} between the reference projections $Y^{(rp)}$ and the bias-corrected model projections $\hat{Y}^{(mp)}$. Panels A and B of \cref{fig:sim_results} show the percentage improvement in the two-dimensional second Wasserstein distance relative to the uncorrected model projections for the inter-variable and cross-location experiments, respectively. In particular, in Panel A, we compute two Wasserstein distances based on both variables and take the average per replicate, while in Panel B, it is based on the two location pairs. All four bias-correction methods improve the distance across the scenarios. As expected, VBC outperforms QM in preserving inter-variable dependence due to its multivariate nature. N-VBC and GN-VBC, meanwhile, perform similarly. However, VBC lacks the ability to preserve spatial dependencies (see Panel B), a limitation that is overcome in both N-VBC and GN-VBC thanks to their nested model structure.
The MSE between the reference and corrected-series autocorrelation functions (ACFs) over lags 1--10 in Panels C and D demonstrates the advantage of the GAM since GN-VBC attains the lowest mean ACF error in both experiments. QM and VBC behave similarly, whereas N-VBC has larger errors that increase as the varied model and reference dependence parameters diverge.

\section{Marginal evaluation}
\label{app:margins}
\begin{figure}
  \centering
  \includegraphics[width=\linewidth]{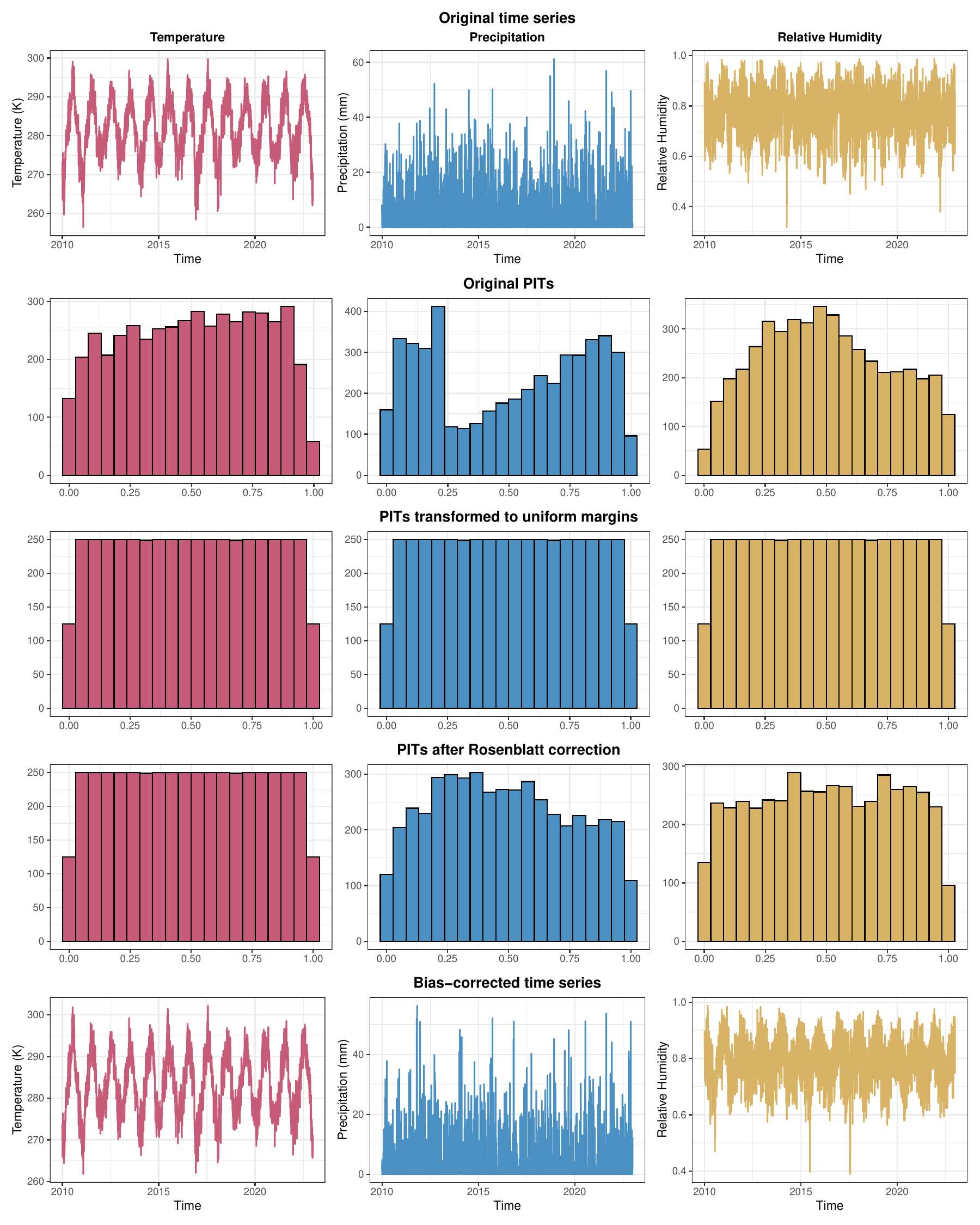}
  \caption{Evaluation of the margins. The top and bottom panels show the time series of the bridging location for temperature (left), precipitation (middle) and relative humidity (right). Panels two to four from the top show histograms of the PIT residuals at different stages of the bias correction pipeline.}
  \label{fig:app:margins}
\end{figure}
To illustrate the behavior of the marginal models used in the vine copula estimation described in \cref{subsec:CUVEE}, \cref{fig:app:margins} displays time series and probability integral transform (PIT) diagnostics at different stages of the proposed GN-VBC pipeline. The results are shown for the selected bridging location (highlighted by a black cross in \cref{fig:map_VD} in the paper) and for three representative variables: near-surface temperature, precipitation, and relative humidity.

The top and bottom panels show the original and bias-corrected time series, respectively. Panels 2 to 4 from the top display histograms of the PIT residuals at successive stages of the marginal processing. In the second panel, PIT values are computed as introduced in \cref{sec:gam}, using the cumulative distribution function implied by the selected parametric family together with the corresponding GAM-based parameter estimates. For temperature and relative humidity, these PITs are already close to uniformly distributed, indicating an adequate marginal fit. In contrast, precipitation, a zero-inflated variable, exhibits a clear deviation from uniformity, reflecting the greater difficulty of capturing its distributional characteristics through a single parametric family.

The GAM-based distributions are used as working models to capture systematic variation and obtain stochastic remainders, rather than being assumed to fully describe the conditional distributions of the data. Deviations from uniformity in the resulting PIT residuals are therefore expected and are addressed by an additional marginal standardization prior to vine copula estimation (third panel). Specifically, each PIT residual $U_{ij}$ introduced in Equation (1) is replaced by
\[
  U^{*}_{ij} = \frac{R_{ij}}{n+1},
\]
where $R_{ij}$ denotes the rank of $U_{ij}$ among $U_{1j},\ldots,U_{nj}$, $n$ is the sample size, and $j$ indexes the variable-location margin, with average ranks assigned to ties. These transformed PITs serve as inputs for the hierarchical vine copulas constructed using the proposed NVC merging strategy. Note that, for the final projection step (i.e.,~the reconstruction of the projected modelled time series), this transformation is reversed so as not to induce any additional noise. Finally, the fourth panel shows the PIT histograms obtained after applying the Rosenblatt and inverse Rosenblatt transforms, illustrating how the dependence correction interacts with the marginal standardization.

Overall, this diagnostic highlights that the marginal processing pipeline successfully produces approximately uniform inputs for copula estimation, even in cases where the initial parametric fit is imperfect. The example of precipitation in particular motivates the use of additional marginal standardization prior to vine copula modelling, thereby supporting the robustness of the proposed framework.

\section{Runtimes}\label{app:runtimes}

Runtimes were measured as wall-clock time within \textsf{R}. All algorithms were executed on the Curnagl computing cluster, each on a single compute node with 20 physical CPU cores and 120\,GB of RAM, allowing for parallel execution where supported by the respective method. Runtime comparisons are based on the original dataset consisting of five variables observed at 22 locations in the canton of Vaud, Switzerland. To ensure comparability across methods, no bootstrap resampling was applied; reported runtimes therefore correspond to a single bias-correction run.
\begin{table}[t!]
  \centering

  \subfloat[\label{app:tab:runtimes_algo}]{%
    \begin{minipage}[t]{0.38\linewidth}
      \centering
      \begin{tabular}{lr}
        \toprule
        QM     &   1.74 \\
        
        \midrule
        MBCn   &  20.29 \\
        R2D2   &   0.80 \\
        \midrule
        VBC    &  80.72 \\
        G-VBC  & 310.43 \\
        \midrule
        C-VBC  & 154.41 \\
        GC-VBC & 483.33 \\
        \midrule
        N-VBC  & 91.01 \\
        GN-VBC & 395.50 \\
        \bottomrule
      \end{tabular}
    \end{minipage}%
  }
  \hfill
  \subfloat[\label{app:tab:runtimes_gam+cuvee}]{%
    \begin{minipage}[t]{0.58\linewidth}
      \centering
      \begin{tabular}{lr}
        \toprule
        GAM derivation                & 129.85 \\
        Compute PITs                  &  18.57 \\
        NVC for $mp$                  &  19.44 \\
        NVC for $rc$                  &  44.90 \\
        Rosenblatt transform          &   0.93 \\
        Inverse Rosenblatt transform  & 1.06 \\
        Inverse CDF from PITs          & 179.51 \\
        \midrule
        Remaining operations          &   1.24 \\
        \bottomrule
      \end{tabular}
    \end{minipage}%
  }

  \caption{Runtimes in seconds for all algorithms compared in the application
  (Panel~a) and step-by-step for our proposed GN-VBC approach (Panel~b).}
  \label{app:tab:runtimes}
\end{table}
\cref{app:tab:runtimes_algo} reports the runtimes (in seconds) of all bias-correction methods considered in \cref{sec:application} (see \cref{tab:methods} in the paper for a description of each algorithm). Empirical QM, which is parallelised across both variables and locations, is approximately twice as slow as R2D2, the fastest algorithm.

Among the methods applied location-wise, namely VBC and G-VBC, incorporating the GAM-based decomposition prior to bias correction substantially increases runtime. Specifically, G-VBC is six times slower than standard VBC, reflecting the computational cost of fitting GAMs independently at each location. When VBC and its GAM-based extension are instead applied jointly to all locations (C-VBC and GC-VBC), this overhead is considerably reduced, with the runtime increasing by a factor of only about~1.6.

Comparing GC-VBC and N-VBC with the baseline C-VBC, both of which are direct extensions of the latter, shows that incorporating NVCs reduces computational costs, whereas the GAM derivation is rather costly. Combining both components in GN-VBC lies in the middle of C-VBC and GC-VBC, indicating that the NVC is more efficient than a standard vine copula; this is also demonstrated by comparing C-VBC and N-VBC.
%Nevertheless, among the approaches that jointly correct all variable-location pairs, R2D2 is the fastest, followed by MBCn. This computational advantage comes at the expense of reduced performance in preserving spatial and/or temporal dependence structures, as demonstrated in Sections 5.3 and 5.4 in the paper. 
Overall, all methods remain computationally feasible, with resource requirements that are reasonable for practical applications.

To further disentangle the computational cost of the proposed GN-VBC approach, \cref{app:tab:runtimes_gam+cuvee} breaks down the total runtime into its main components, corresponding to the steps described in \cref{sec:gnvbc} in the paper. The most time-consuming operation is the last step of the pipeline, the back-transformation via inverse marginal CDFs, followed by the GAM fitting step. The construction of the two NVCs accounts for approximately $16\%$ of the total runtime. All remaining data transformations and auxiliary computations are grouped under ``remaining operations'' and contribute only marginally to the overall computational cost.

\section{Results in Tables} \label{app:tables}
\cref{tab:inter-var} shows the exact values of improvement in second Wasserstein distances depicted in \cref{fig:wd_loc} in the paper rounded to two decimals. \cref{tab:wd} and \cref{tab:mse} show the corresponding tables for \cref{fig:spat} (spatial consistency) based on 20 bootstrap samples, while \cref{tab:acf_mean} shows mean and standard deviations in 365-day ACF differences (temporal consistency) corresponding to \cref{fig:acf_365} in the paper.

\begin{table}[ht]
  \centering
  \begin{tabular}{r|ccccccccc}
    \hline
    Id & QM & MBCn & R2D2 & VBC & G-VBC & C-VBC & GC-VBC & N-VBC & GN-VBC \\
    \hline
    30  & 76.07 & 80.64 & 79.52 & 80.36 & 79.99 & 73.25 & 76.70 & 78.10 & 79.71 \\
    31  & 84.30 & 86.73 & 86.20 & 86.09 & 85.82 & 80.46 & 83.77 & 85.34 & 86.30 \\
    32  & 89.81 & 91.65 & 91.60 & 91.32 & 90.98 & 87.66 & 88.89 & 91.31 & 91.31 \\
    52  & 74.11 & 79.93 & 79.04 & 78.96 & 78.34 & 69.43 & 75.68 & 78.49 & 78.46 \\
    53  & 80.51 & 84.14 & 83.83 & 83.57 & 83.31 & 77.04 & 80.47 & 83.73 & 83.71 \\
    54  & 87.89 & 89.67 & 89.62 & 89.28 & 89.09 & 85.67 & 86.59 & 89.34 & 89.39 \\
    73  & 85.67 & 89.36 & 88.63 & 88.83 & 88.68 & 83.76 & 87.09 & 88.96 & 88.29 \\
    74  & 75.37 & 81.04 & 80.42 & 80.56 & 79.46 & 71.53 & 77.51 & 80.57 & 80.22 \\
    75  & 76.05 & 80.47 & 80.02 & 79.26 & 78.53 & 70.90 & 75.90 & 79.76 & 79.78 \\
    94  & 85.26 & 87.90 & 87.21 & 87.55 & 85.86 & 80.80 & 84.65 & 87.49 & 86.48 \\
    95  & 86.72 & 89.24 & 88.88 & 88.99 & 88.59 & 83.05 & 86.49 & 89.03 & 88.37 \\
    96  & 68.20 & 73.96 & 73.24 & 73.29 & 71.72 & 60.45 & 67.85 & 71.88 & 73.30 \\
    97  & 80.33 & 83.97 & 83.59 & 82.89 & 82.54 & 76.29 & 79.92 & 81.97 & 83.42 \\
    115 & 83.76 & 86.67 & 86.39 & 85.84 & 84.88 & 79.91 & 83.13 & 85.62 & 84.60 \\
    116 & 87.11 & 89.12 & 88.96 & 88.96 & 87.89 & 83.69 & 86.68 & 88.55 & 87.91 \\
    117 & 82.30 & 85.05 & 84.50 & 84.80 & 83.98 & 77.94 & 81.49 & 84.06 & 84.00 \\
    118 & 52.94 & 61.74 & 59.98 & 61.33 & 59.54 & 43.52 & 52.32 & 56.14 & 58.81 \\
    119 & 78.01 & 81.56 & 80.67 & 80.91 & 80.18 & 73.46 & 76.82 & 79.94 & 81.00 \\
    134 & 88.37 & 91.03 & 90.82 & 90.97 & 89.32 & 89.81 & 88.40 & 89.67 & 89.19 \\
    135 & 90.55 & 92.43 & 92.58 & 92.25 & 90.61 & 90.95 & 90.45 & 91.54 & 90.99 \\
    155 & 82.33 & 85.73 & 85.80 & 85.40 & 83.35 & 82.57 & 81.80 & 83.13 & 82.97 \\
    156 & 78.86 & 82.37 & 82.45 & 81.78 & 79.91 & 78.42 & 77.85 & 79.41 & 78.70 \\
    \hline
  \end{tabular}
  \caption{Preservation of inter-variable dependencies: Improvement in second Wasserstein distance for each location and algorithm. The corresponding ids are depicted in \cref{app:map_ids}.}
  \label{tab:inter-var}
\end{table}

\begin{figure}
  \centering
  \includegraphics[width=0.6\linewidth]{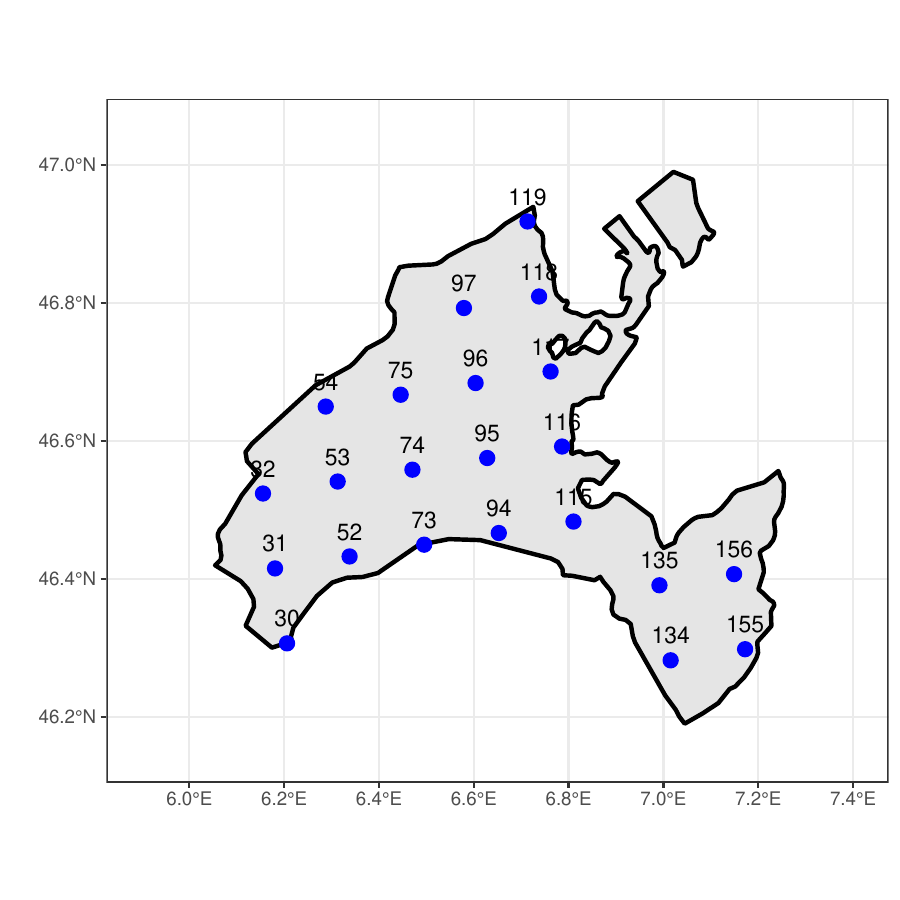}
  \caption{The grid cells with corresponding ids in the canton of Vaud, Switzerland.}
  \label{app:map_ids}
\end{figure}

\begin{table}[ht]
  \centering
  \begin{tabular}{lccccc}
    \hline
    Method & tas & pr & hurs & sfcWind & ps \\
    \hline
    QM     &
    59.36 (6.31) & 20.84 (5.31) & 4.45 (6.81) & 80.91 (0.52) & 95.26 (0.72) \\

    MBCn   &
    48.12 (4.81) & 21.27 (5.81) & 20.63 (8.04) & 84.98 (0.59) & 94.63 (0.64) \\

    R2D2   &
    62.18 (7.15) & 25.05 (6.16) & 33.03 (11.61) & 89.72 (1.21) & 95.39 (0.88) \\

    VBC    &
    -0.74 (10.29) & 23.74 (3.94) & -11.97 (6.69) & 75.90 (3.65) & 95.25 (0.75) \\

    C-VBC  &
    47.02 (5.93) & 16.22 (5.28) & 17.37 (8.68) & 84.61 (5.55) & 94.35 (0.56) \\

    G-VBC  &
    54.92 (4.59) & 18.51 (6.59) & -4.39 (6.58) & 77.60 (0.70) & 95.76 (0.80) \\

    GC-VBC &
    52.54 (6.70) & 15.72 (6.77) & 19.92 (9.40) & 84.55 (0.69) & 94.68 (0.52) \\

    N-VBC  &
    50.55 (5.05) & 7.79 (6.46) & 24.76 (9.76) & 88.20 (0.51) & 94.32 (0.62) \\

    GN-VBC &
    63.10 (5.74) & 29.94 (7.72) & 27.25 (11.68) & 86.30 (0.75) & 93.95 (1.14) \\
    \hline
  \end{tabular}
  \caption{Preservation of spatial dependencies: Mean and standard deviations of improvements in second Wasserstein distances for each method and variable.}
  \label{tab:wd}
\end{table}

\begin{table}[ht]
  \centering
  \begin{tabular}{lccccc}
    \hline
    Method & tas & pr & hurs & sfcWind & ps \\
    \hline
    Model  &
    1.08 (0.74) &
    16.25 (29.27) &
    37.09 (37.19) &
    53.03 (45.59) &
    157.41 (217.10) \\

    QM     &
    0.90 (0.68) &
    10.22 (12.16) &
    35.59 (35.58) &
    56.80 (46.96) &
    103.74 (147.78) \\

    MBCn   &
    20.77 (25.23) &
    21.28 (22.36) &
    20.82 (18.93) &
    19.08 (12.58) &
    44.92 (54.44) \\

    R2D2   &
    0.09 (0.12) &
    8.54 (9.17) &
    1.13 (1.99) &
    0.15 (0.32) &
    8.74 (13.20) \\

    VBC    &
    4.59 (5.43) &
    11.11 (11.51) &
    46.52 (41.25) &
    72.27 (56.54) &
    109.62 (155.01) \\

    C-VBC  &
    2.12 (2.90) &
    11.80 (13.61) &
    7.61 (9.29) &
    8.57 (12.60) &
    21.15 (35.91) \\

    G-VBC  &
    3.79 (4.94) &
    11.19 (11.19) &
    33.63 (30.69) &
    58.80 (46.50) &
    9.21 (11.34) \\

    GC-VBC &
    0.90 (1.28) &
    9.97 (10.09) &
    9.21 (13.29) &
    9.33 (13.07) &
    22.93 (39.23) \\

    N-VBC  &
    1.83 (1.77) &
    9.13 (10.32) &
    5.79 (7.62) &
    4.78 (6.42) &
    13.09 (19.99) \\

    GN-VBC &
    0.55 (0.76) &
    7.40 (11.46) &
    1.10 (1.77) &
    2.97 (4.31) &
    7.75 (17.11) \\
    \hline
  \end{tabular}
  \caption{Preservation of spatial dependencies: Mean and standard deviations of MSE of spatial correlation between the reference and all considered methods for all variables. Values are multiplied by 1000 and rounded to two digits.}
  \label{tab:mse}
\end{table}

\begin{table}[ht]
  \centering
  \begin{tabular}{lccccc}
    \hline
    Method & tas & pr & hurs & sfcWind & ps \\
    \hline
    Model  &
    1.57 (2.67) &
    0.42 (0.58) &
    12.63 (16.02) &
    4.23 (8.33) &
    2.43 (3.37) \\

    QM     &
    0.89 (1.32) &
    0.41 (0.57) &
    12.54 (15.73) &
    4.36 (8.35) &
    2.24 (3.12) \\

    MBCn   &
    2.00 (1.93) &
    0.41 (0.56) &
    4.94 (6.64) &
    0.61 (0.87) &
    1.68 (2.39) \\

    R2D2   &
    0.33 (0.44) &
    0.56 (2.75) &
    9.44 (17.90) &
    1.42 (9.01) &
    5.13 (37.82) \\

    VBC    &
    16.88 (23.61) &
    0.42 (0.61) &
    11.10 (14.50) &
    4.48 (9.84) &
    2.27 (3.15) \\

    C-VBC  &
    148.56 (108.87) &
    0.66 (0.94) &
    11.47 (12.08) &
    6.50 (10.79) &
    1.82 (3.00) \\

    G-VBC  &
    0.18 (0.32) &
    0.40 (0.55) &
    2.38 (3.25) &
    0.78 (1.15) &
    2.01 (2.84) \\

    GC-VBC &
    0.61 (0.76) &
    0.45 (0.64) &
    4.65 (7.73) &
    1.49 (3.24) &
    2.08 (2.99) \\

    N-VBC  &
    1.13 (1.70) &
    0.52 (0.72) &
    2.14 (3.55) &
    1.24 (2.66) &
    1.81 (2.54) \\

    GN-VBC &
    0.25 (0.46) &
    0.40 (0.56) &
    1.13 (1.51) &
    0.83 (1.21) &
    2.29 (3.48) \\
    \hline
  \end{tabular}
  \caption{Preservation of temporal dependence: Mean and standard deviation over 365 days of differences in ACF per method and variable. Values are multiplied by 1000 and rounded to two digits.}
  \label{tab:acf_mean}
\end{table}